\documentclass[preprintnumbers,11pt,nofootinbib]{revtex4-1}
\usepackage{amsmath,amsfonts,bm}
\usepackage{graphicx}
\usepackage{color}
\usepackage{subfigure}

\usepackage{verbatim}

\begin{document}
\title{Adsorption of solutes at liquid-vapor
    interfaces: Insights from lattice gas models}
\author{Suriyanarayanan Vaikuntanathan,\textit{$^{1}$} Patrick R. Shaffer,\textit{$^{2}$} and
Phillip L. Geissler\textit{$^{1,2}$}}
\affiliation{$^{1}$~Chemical Sciences Division, Lawrence Berkeley National Lab, Berkeley, CA 94720\\ 
$^{2}$~Department of Chemistry, University of California, Berkeley, CA 94720}
\begin{abstract}
The adsorption behavior of ions at liquid-vapor interfaces exhibits
several unexpected yet generic features. In particular, energy and
entropy are both minimum when the solute resides near the surface, for
a variety of ions in a range of polar solvents, contrary to
predictions of classical theories. Motivated by this generality, and
by the simple physical ingredients implicated by computational
studies, we have examined interfacial solvation in highly schematic
models, which resolve only coarse fluctuations in solvent density and
cohesive energy. Here we show that even such lattice gas models
recapitulate surprising thermodynamic trends observed in detailed
simulations and experiments. Attention is focused on the case of two
dimensions, for which approximate energy and entropy profiles can be
calculated analytically. Simulations and theoretical analysis of the
lattice gas highlight the role of capillary wave-like fluctuations in
mediating adsorption. They further point to ranges of temperature and
solute-solvent interaction strength where surface propensity is
expected to be strongest.

\end{abstract}
\maketitle 

\section{Introduction}
Experiments and computer simulations have demonstrated that certain
small ions preferentially adsorb at liquid-vapor
interfaces~\cite{Jungwirth:2006,Jungwirth:2002,Petersen:2004,NoahVanhoucke:2009,Otten:2012},
contrary to expectations from classic theories of ion
solvation~\cite{Onsager:1934}.  Efforts to explain this behavior have
focused primarily on accounting for effects of solute polarizability
and the thermodynamic cost of excluding volume in bulk
solution~\cite{Jungwirth:2006,Archontis:2006,Levin:2009,
  Levin:2009b,Markin:2002}. For small excluded volumes, this cost is primarily entropic and reflects constraints on the available arrangements of nearby solvent
molecules.~\cite{Huang:2002}. From this perspective the adsorption of a
non-polarizable ion is expected to be opposed energetically (through
loss of dielectric polarization energy) and favored entropically
(through recovered freedom of local solvent arrangements).  Very
recent work has shown, however, that these thermodynamic driving
forces generally follow the opposite trend, with energy and entropy
both exhibiting minima when an ion resides near the
interface~\cite{Otten:2012,Caleman:2011,Iuchi:2009}.  In these studies we and
others also revealed important driving forces governing ion adsorption
that had been largely ignored in discussions of interfacial solvation.

The interface between a dense polar liquid and its coexisting vapor is
a region of high energy density (relative to that of the bulk liquid),
due to incomplete or strained coordination of solvent molecules. The
liquid's boundary can nonetheless be quite soft, with substantial
wavelike shape fluctuations even on molecular length scales. These
generic facts appear from simulations to have important consequences
for the thermodynamics of ion adsorption. In particular, moving a
solute from bulk solution to the interface effects more than reduced
coordination of the ion (which is exclusively emphasized in classic
theories). An adsorbed solute also occupies volume in the high-energy
region, effectively returning solvent density to the bulk
environment~\cite{Stuart:1996,Vaitheeswaran:2006,Otten:2012}. The consequence of this simple
but often overlooked accounting is a substantially favorable energetic
contribution. For ions of modest size it is sufficient to render
adsorption energies negative~\cite{Stuart:1996,Vaitheeswaran:2006,Otten:2012,Caleman:2011,Iuchi:2009}. Although electrostatic forces clearly
figure prominently in this mechanism, their long spatial range is not
manifested by the dominant energetic changes accompanying
adsorption~\cite{Otten:2012}. Instead, a spatially local approximation, which expresses
the total energy in terms of the populations of different solvent
regions (bulk, interface, and solute coordination zone),
\begin{equation}\label{eq:Ulocal}U_{local}=\epsilon_{bulk}n_{bulk}+\epsilon_{interface}n_{interface}+\epsilon_{coord}n_{coord}\,,
\end{equation}
was found to be quite accurate for systems with non-polarizable ions~\cite{Otten:2012}. Here, $\epsilon_\gamma$ is the energy
per molecule of solvent in region $\gamma$, and $n_\gamma$ is the
corresponding number of molecules in that region.

Softness of the interface, also neglected by conventional theories, is
implicated by these simulations to significantly influence the entropy
of adsorption~\cite{Otten:2012}. A solute that substantially attracts solvent, such as
an ion in water, tends to suppress surface undulations when it resides
near the interface. The entropic cost of this suppression can be
considerable, opposing adsorption by many $k_{\rm B}T$.  Harmonic
analysis of capillary wave-like fluctuations in the adsorbed and
non-adsorbed states was found to roughly account for observed entropy
profiles~\cite{Otten:2012}.

The two physical ingredients speculated in Ref.~\cite{Otten:2012} to
be key for ion adsorption, namely surface shape fluctuations and
heterogeneous energy density near the liquid's boundary, are so
generic as to be described by the very simplest microscopic models of
the liquid state.  Motivated by this apparent generality, we examine
in this paper the classic reduced model of liquid-vapor coexistence,
suitably adapted to address questions of solvation. Specifically, we
consider a two-dimensional lattice gas, with a solute that occupies a
single lattice cell and attracts only its nearest neighbors.
Presenting results of both theory and simulation, we demonstrate that
the unexpected thermodynamic trends revealed by recent atomistic
simulations can indeed be captured by such a simple model.  The very
limited microscopic detail in the case of a lattice gas allows
unambiguous conclusions regarding the source of energy and entropy
changes as the solute moves through the interface. It allows as well
crude predictions for the varying propensity ions exhibit for
liquid-vapor interfaces.

The rest of the paper is organized as follows. In
Sec.~\ref{sec:Models}, we describe our lattice gas model for
solvation. We present results of Monte Carlo simulations in
Sec.~\ref{sec:MCS}. In Sec.~\ref{sec:SOS} we develop a theory for the
behavior of this model based on the so-called solid-on-solid
simplification and compare its predictions with numerical results.
We conclude in Sec.~\ref{sec:Conclusion} with a discussion of specific implications
for three-dimensional molecular systems.

\begin{figure}[tbp]
\begin{center}
   \subfigure[]{
   \label{fig:Ep15}
   \includegraphics[scale=0.25,angle=0]{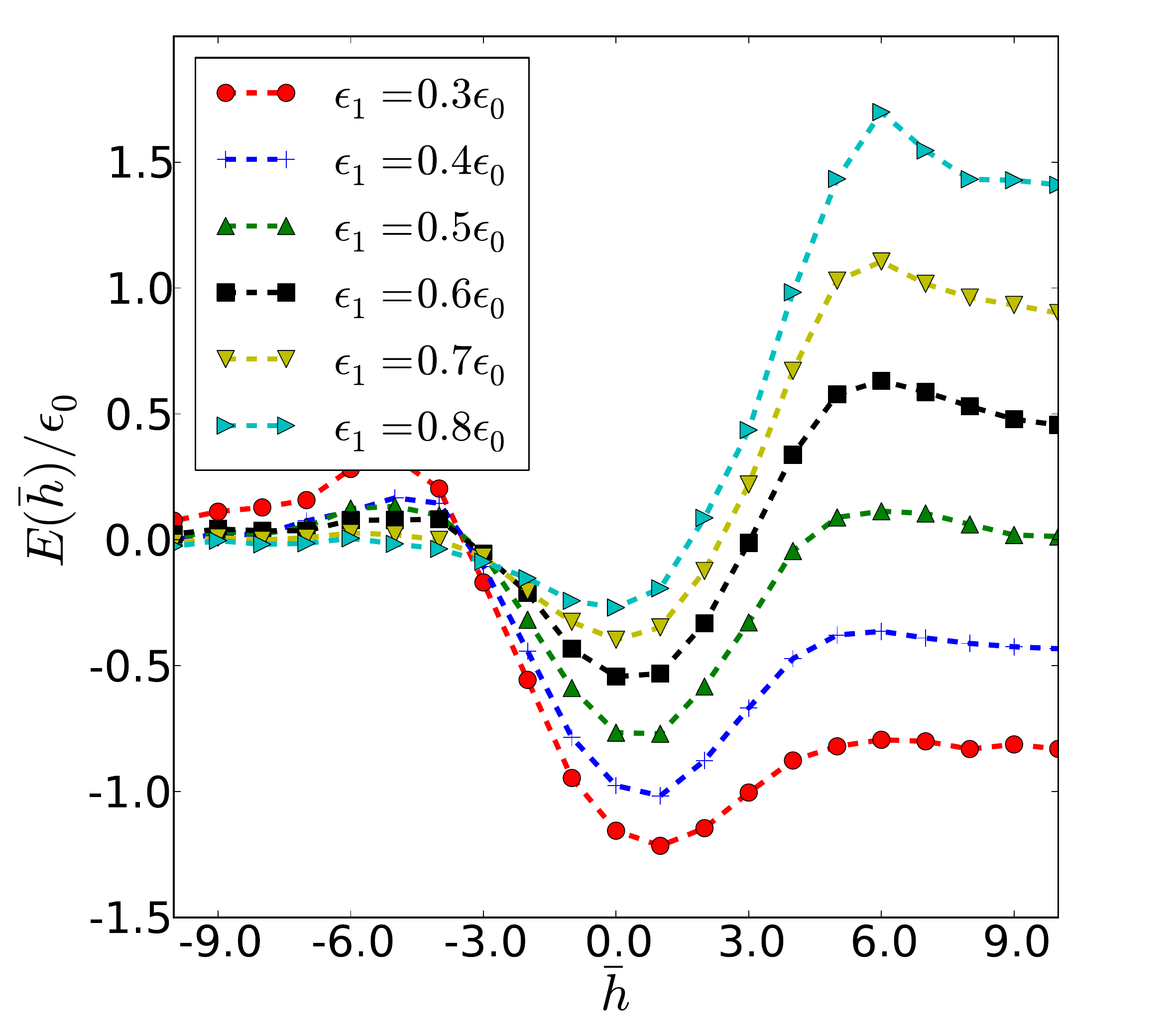}
   }
     \subfigure[]{
   \label{fig:Ep18}
   \includegraphics[scale=0.25,angle=0]{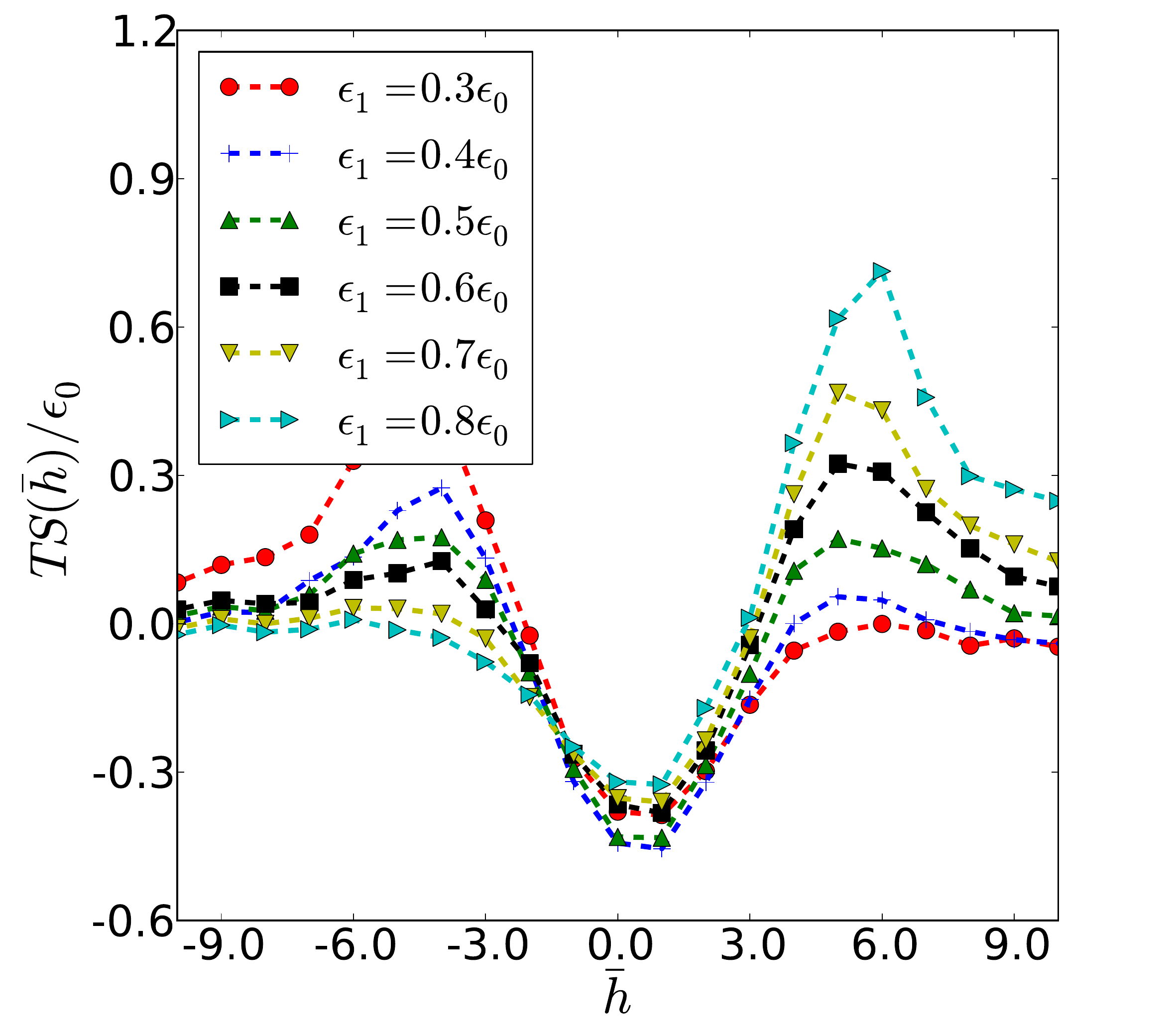}
   }
   \subfigure[]{
   \label{fig:Ep21}
   \includegraphics[scale=0.25,angle=0]{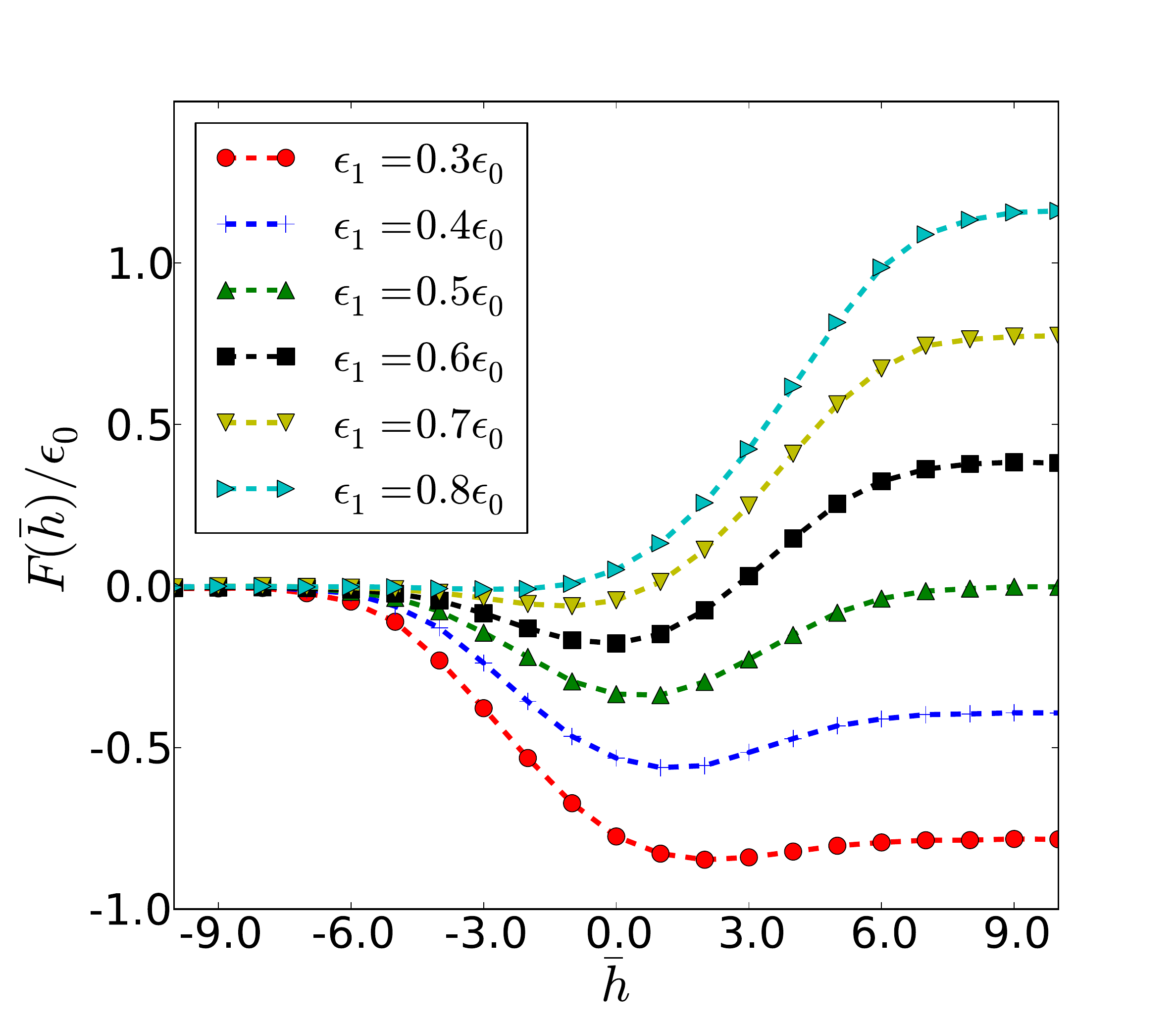}
   }
   
   \end{center}
  \caption{Profiles of (a) energy $E(\bar{h})$, (b) entropy $S(\bar{h})$, and (c) free energy $F(\bar{h})$, as a function of the height of the solute, $\bar h$, for various values of $\epsilon_1$ at $T=\epsilon_0/3k_B$. The height of $\bar h=0$ coincides with the GDS. These profiles were obtained from Monte Carlo simulations of the lattice gas.}
\label{fig:1DprofilesEp3}
\end{figure}

\section{Lattice gas model}
\label{sec:Models}
We consider a two-dimensional $L\times L$ lattice gas system with 
a single solute, that occupies one lattice cell. Cells are indexed by a horizontal
coordinate $x$ ranging from 1 to $L$ and a vertical coordinate $y$
ranging from $-(L-1)/2$ to $(L-1)/2$~\footnote{Without loss of
  generality, we assume that $L$ is odd.}. The x-coordinate of the solute is constrained to $x=(L+1)/2$. The solute interacts only
with the 4 nearest neighbor cells. It additionally excludes volume, so
that a cell may not be simultaneously occupied by both solvent and
solute. This system is described by the Hamiltonian
\begin{equation}
\label{eq:latticegas}
H=H_0-\epsilon_1\sum_{i,j\in nn} n^s_in_j,
\end{equation}
where
\begin{equation}
\label{eq:latticegas2}
H_0\equiv-{\epsilon_0}\sum_{i,j\in nn} n_in_j-\mu\sum_{i}n_i
\end{equation}
describes the pure solvent. The occupation variables $n$ and $n_i^s$
are binary:
$n_i=1$ if the $i^{th}$ lattice site is occupied by a solvent
and zero otherwise, $n^s_i=1$ if the $i^{th}$ lattice site is occupied
by a solute and zero otherwise. $\sum_{i,j\in nn}$ denotes a sum
over all nearest neighbor lattice pairs $i$, $j$. The coupling
constant $\epsilon_0 > 0$ describes attraction between neighboring solvent
cells;  
$\epsilon_1$ is the solvent-solute coupling constant.  

For both theory and simulation we work with a grand canonical ensemble,
in which the number of solvent cells can fluctuate subject to chemical
potential $\mu$. (The number of solute cells is fixed.)
Conditions of
liquid-vapor coexistence are imposed by setting $\mu=-2\epsilon_0$,
and applying boundary conditions that favor liquid in $y<0$ and vapor
in $y>0$.  Specifically, lattice cells with $y=-(L-1)/2$ are constrained
to be occupied by solvent; those with $y=(L-1)/2$ are constrained to be
unoccupied (i.e., vapor).
  
The mean interface position, or the Gibbs Dividing Surface (GDS), is
fixed at $y=0$ by constraining as well the state of lattice cells with
$x=1$ and $x=L$; those with $y<0$ are occupied, and those with $y>0$
are unoccupied.

We will examine values of $\epsilon_1$ between 0 and 1. By symmetry
(see Appendix A), a solute with $\epsilon_1 = \epsilon_0/2$ is equally
likely to reside in a cell well within the liquid phase as in a cell
well within the vapor phase. Judging by the free energy of transfer
from liquid to vapor, one would therefore consider solutes with
$\epsilon_1 < \epsilon_0/2$ to be solvophobic, and those with
$\epsilon_1 > \epsilon_0/2$ to be solvophilic (although the latter may
attract solvent significantly less strongly than adjacent solvent
cells attract each other).

Using theory and simulation, we have determined the
free energy $F$, average energy $E$, and the entropy $S$
as functions of the height $\bar{h}$ of the solute (i.e., its $y$-coordinate).

\section{Monte Carlo Simulations}
\label{sec:MCS}
We present results of Monte Carlo simulations at temperature
$T=\epsilon_0/3 k_{\rm B}$ (well below the critical temperature $T_c
\approx \epsilon_0 /1.764k_{\rm B}$) for a lattice with linear
dimension $L=30$.
Trial moves included (i) changing the state of a cell from solvent to
vapor, (ii) changing a cell from vapor to solvent, and (iii)
translating the solute along the $y$ axis from a lattice cell $s$ to
a cell $s'$. In the case of solute displacements, two
different trial moves were used, distinguished by their treatment of
solvent occupation in the cell $s$ vacated by the solute. In one move,
cell $s$ adopts the previous state of $s'$, i.e., $n_s^{trial} =
n_{s'}$. In another move, cell $s$ adopts the complementary state of
$s'$, i.e., $n_s^{trial} = 1-n_{s'}$. In all cases detailed balance
was ensured through appropriate Metropolis acceptance criteria,
preserving the Boltzmann distribution $p(\{n_i, n_i^s  \}) \propto
e^{-\beta H}$, where $\beta = 1/k_{\rm B}T$ is inverse temperature.

We obtained estimates of $F(\bar h)$ by computing
the equilibrium probability $p(\bar h)\propto \exp[-\beta F(\bar
h)]$ that the solute resides at height $\bar h$. The average energy
$E(\bar h)$ was estimated directly from simulations, while estimates
of the entropy $S(\bar h)$ were calculated from $TS(\bar h)=E(\bar
h)-F(\bar h)$.

The profiles of $F(\bar h)$, $E(\bar h)$, and $S(\bar h)$ obtained
from simulations with
$\epsilon_1/\epsilon_0=\{0.3,0.4,0.5,0.6,0.7,0.8\}$, are plotted in
Figs.~\ref{fig:1DprofilesEp3}.  These results echo several nontrivial
thermodynamic trends observed in molecular simulations. In particular,
free energy profiles for
$\epsilon_1=\{0.4\epsilon_0,0.5\epsilon_0,0.6\epsilon_0,0.7\epsilon_0\}$
exhibit minima near $\bar{h}=0$. These solutes adsorb to the
interface, i.e., their density at $\bar{h}=0$ is higher than in either
bulk phase.  Entropy minima at the same values of $\epsilon_1$
indicate that this adsorption reduces the diversity of accessible
solvent configurations. The driving force for surface propensity is
instead energetic, as evidenced by strong minima in $E(\bar h)$ near
$\bar{h}=0$. Interestingly, adsorption is strongest for the
case $\epsilon_1=\epsilon_0/2$, at the crossover between solvophobic
and solvophilic regimes. 

These trends were also observed in simulations with smaller and larger
system sizes ($L=\{20,30,40,50,60,70\}$), and for various temperatures
in the range $\epsilon_0/5k_B\leq T\leq \epsilon_0/2k_B$. In all
cases, profiles of free energy, energy, and entropy exhibit minima
when the solute is at the interface, and maximum adsorption occurs at
$\epsilon_1=\epsilon_0/2$.

\begin{figure}[h!]
\begin{center}
   \subfigure[]{
   \label{fig:Zin}
   \includegraphics[scale=0.23,angle=0]{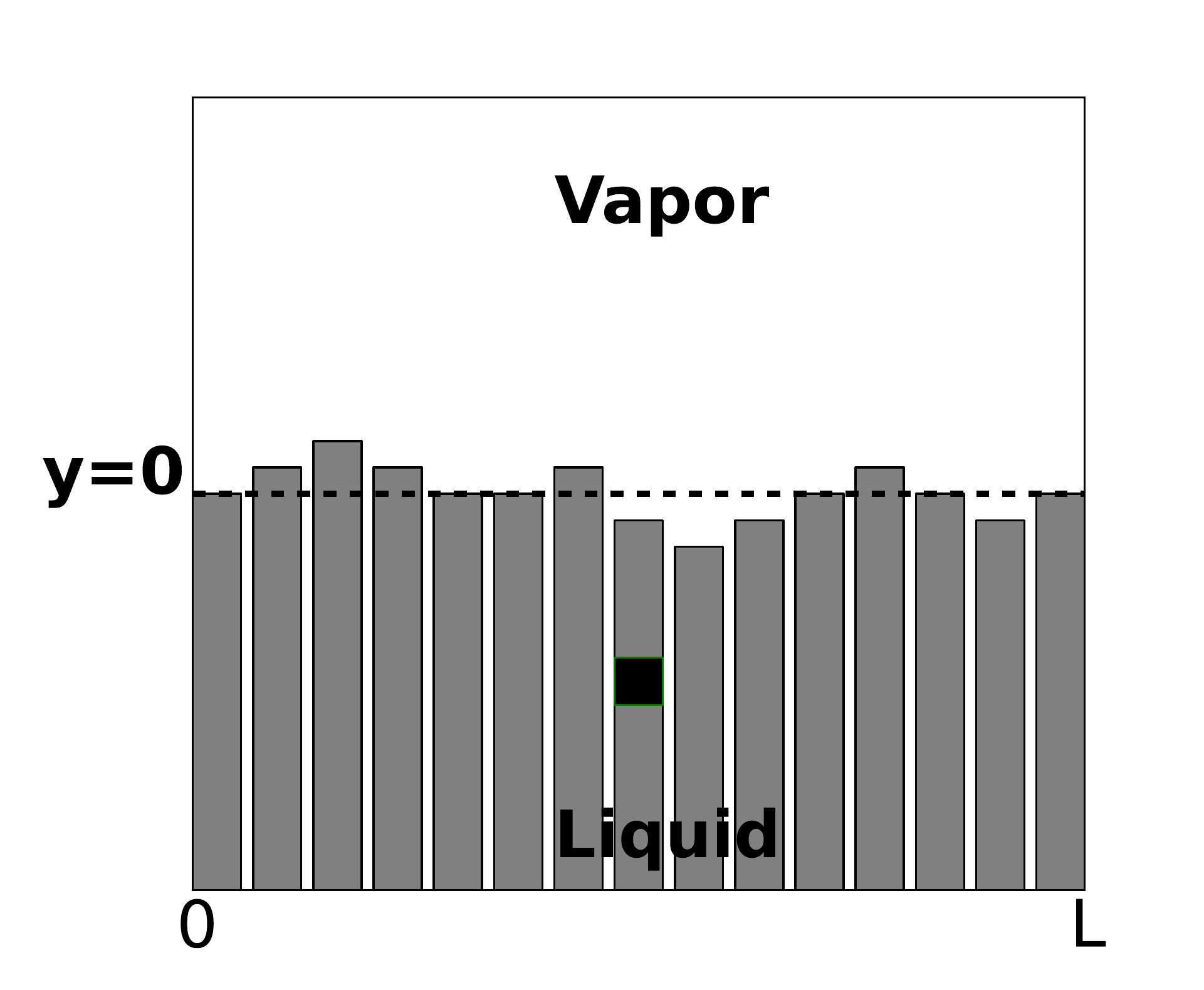}
   }
     \subfigure[]{
   \label{fig:Zpinned}
   \includegraphics[scale=0.23,angle=0]{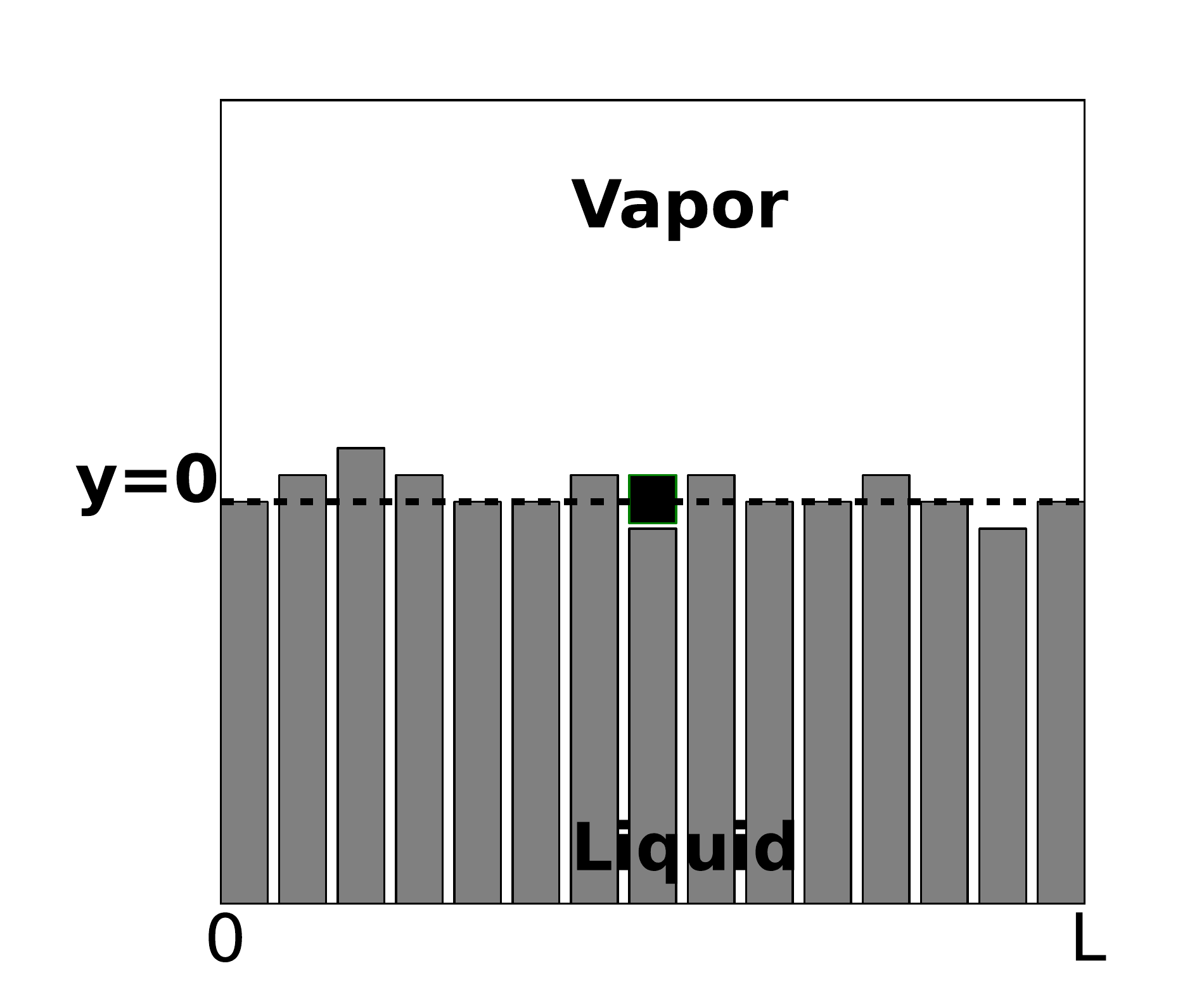}
   }
   \subfigure[]{
   \label{fig:Zout}
   \includegraphics[scale=0.23,angle=0]{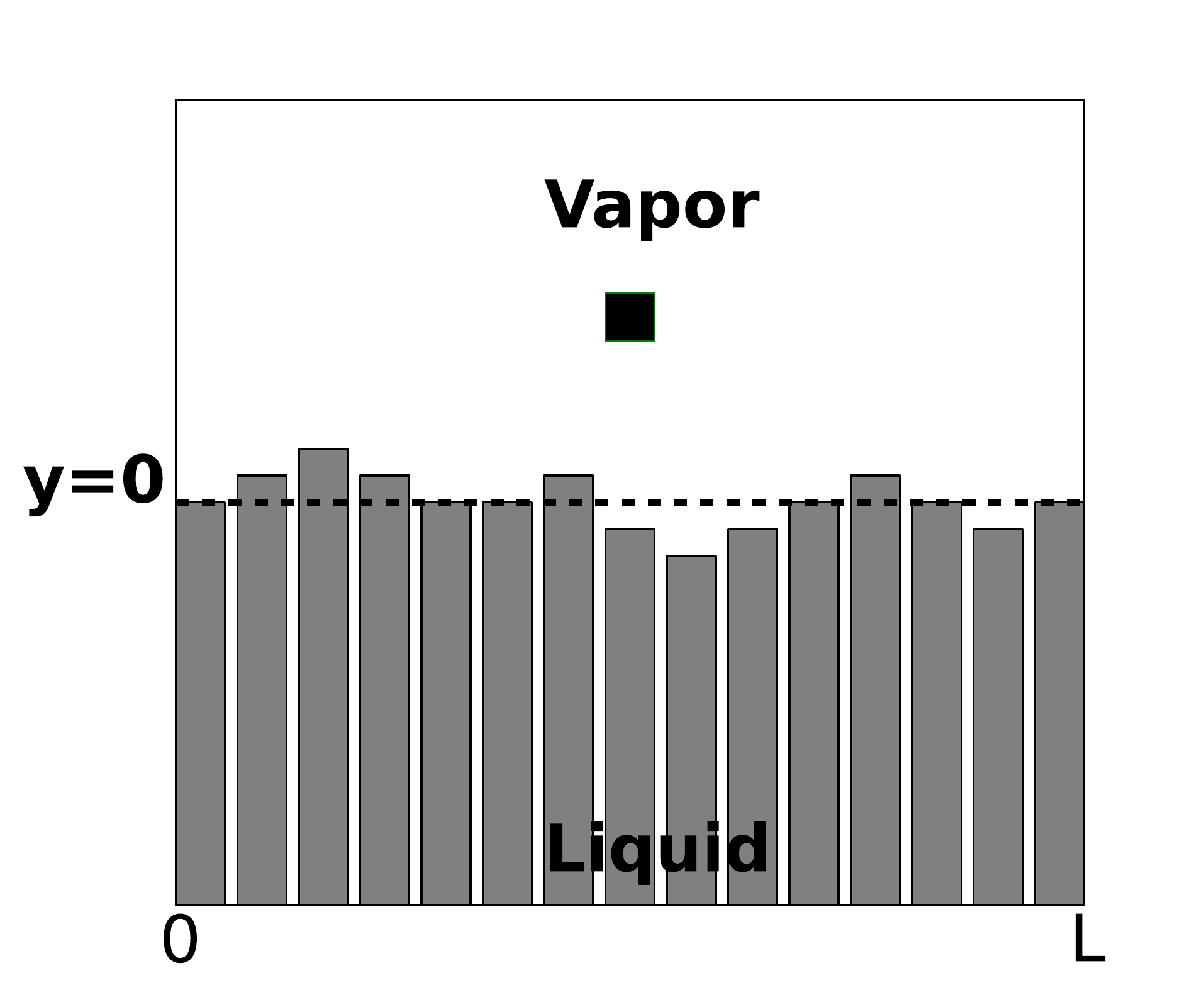}
   }

  \caption{Illustration of three configurational subensembles in the RSOS model. The fully solvated \textit{in} state (a) features a liquid vapor interface entirely above the solute. In the \textit{pinned} state subensemble (b), this interface directly contacts the solute. In the unsolvated state \textit{out} state, (c), the interface lies below the solute.}
\label{fig:latticegas}
\end{center}
\end{figure}

\section{Theory}
\label{sec:SOS}

The 2-d lattice gas model of solvation described by Eq.~\ref{eq:latticegas}, within
approximations that are well justified for $T \ll T_c$, is
sufficiently simple to permit an analytical solution. This section
presents our mathematical analysis, which draws heavily on results
from theoretical studies of surface roughening in Ising-like models.
\cite{Weeks:1977,Chui:1981,Abraham:1980,Fisher:1984}
Our key assumptions are that the liquid-vapor interface is well
defined and that the solute does not significantly alter fluctuations
within either phase. More specifically, we restrict attention to the
so-called ``solid-on-solid'' (SOS) limit~\cite{Nelson:2004,Temperley:1952}, in which: (1) all cells in the
liquid phase are occupied by solvent (except the cell occupied by the
solute when it is present), (2) all cells in the vapor phase are
unoccupied (except the solute cell), and (3) the boundary between
liquid and vapor is a single-valued function of $x$. With this neglect
of bubbles, islands, and overhangs, a configuration can be specified
through the heights of $L$ columns representing the instantaneous 
domain of the liquid phase.

Under conditions of coexistence, the Hamiltonian describing the solvent 
is then given by~\cite{Nelson:2004,Safran:2003} 
\begin{equation}
\label{eq:barehamil}
\bar{H_0}= \frac{\epsilon_0}{2} \sum_{i,j \in nn}|h_i-h_{j}|  \,,
\end{equation}
where $h_i$ denotes the height of the liquid in the $i^{th}$ column.
Boundary conditions described in Sec.~\ref{sec:Models} fix the heights of the first
and last columns, $h_1=0$ and $h_L=0$. They also bound the range of
column height fluctuations, $-L/2 < h_i < L/2$. As an added simplification, we work with the Restricted SOS (RSOS)
model{~\cite{Chui:1981} where the height difference between
  neighboring columns is constrained to be at most 1,
  $|h_i-h_{i+1}|\leq1$. A few allowed configurations of the RSOS
model are illustrated in Fig~\ref{fig:latticegas}.
Previous work has shown that fluctuations of the neat liquid-vapor interface
of a two-dimensional lattice gas are well described by the
RSOS model for temperatures $T \ll T_c$~\cite{Chui:1981,Fisher:1984}.
As in that work, the restriction on height fluctuations is necessary to
derive simple closed form expressions for
thermodynamic quantities of interest.

The liquid-vapor interface of the 
RSOS model supports shape fluctuations akin to capillary waves.  It
has been shown in particular that long-wavelength Fourier modes are
Gaussian distributed to a good approximation. For a 2-d
RSOS interface of length $M$, whose boundaries are fixed at $h_1=p$
and $h_M=q$, the partition function is given by
~\cite{Chui:1981,Fisher:1984,Kardar2007}
\begin{equation}
 \label{eq:ZM}
Z^{1,M}(p,q)\approx \frac{e^{2aM}}{\sqrt{4\pi aM}}\exp\left[{\frac{-(p-q)^2}{4aM}}\right]\,,
\end{equation}
in the limit $Ma>>1$, where $a\equiv \exp(-\beta\epsilon_0/2)$.
This result is strictly valid only in the low temperature limit, where
terms of order $O(a^2)$ can be neglected. 
Empirically, however,
the thermodynamic properties we compute using Eq.~\ref{eq:ZM} agree 
well with those obtained from lattice gas
simulations even at moderate temperatures.
 
To construct a theory for interfacial solvation from Eq.~\ref{eq:ZM},
we first partition the ensemble of solvent fluctuations consistent
with a given solute height $\bar{h}$. We classify configurations in
which the interface lies entirely above the solute as belonging to a
completely solvated class, denoted \textit{in}. Immersion in this sense
requires $h_s > \bar{h}$, where $s$ is the solute's $x$-coordinate.
Configurations in which the solute makes no contact with the liquid
phase, $h_s < \bar{h}-1$, we classify as \textit{out}. In the remaining
configurations, which we classify as \textit{pinned}, the solute resides at
the interface, $h_s = \bar{h}-1$, in effect pinning local height
fluctuations.

Defining $Z_{in}(\bar{h})$, $Z_{out}(\bar{h})$, and
$Z_{pinned}(\bar{h})$ as the partition functions for these three
subensembles, we can reconstruct the total partition function for a
given solute height simply as
\begin{equation}
\label{eq:Zhs}
Z(\bar{h}) = Z_{in}(\bar h) + Z_{out} (\bar h) +Z_{pinned}(\bar h)\,.
\end{equation}
Each of the three configurational classes corresponds to a constrained
ensemble of pure liquid-vapor configurations, reweighted by
solvent-solute interactions. We will calculate their partition
functions in turn.

The \textit{out} subensemble is simplest,
since solvent and solute do not interact. The partition function for
this case is given by
\begin{equation}
Z_{out}(\bar{h}) = \mathop{\sum\dots\sum}_{h_1=0\,,h_L=0\,,h_s < \bar{h}-1} e^{-\beta \bar H_0}=
\sum_{h_{s}<\bar h-1}Z^{1,s}(0,h_{s})Z^{s,L}(h_{s},0)\,,
\end{equation}
where the notation $\mathop{\sum\dots\sum}_{h_1=0\,,h_L=0\,,h_s < \bar{h}-1}$ implies a sum over all configurations of the RSOS system subject to the constraints $\{h_1=0\,,h_L=0\,,h_s < \bar{h}-1\}$. 
To obtain a closed-form expression, we convert sums to integrals in the limit $L>>1$, and set the solute's horizontal position at $s=(L+1)/2$, 
yielding,
\begin{equation}
\label{eq:Zhs3}
\begin{split}
Z_{out}(\bar h)&\approx \frac{e^{2aL}}{2\pi aL}\int _{-(L-1)/2}^{\bar h-1}dh_se^{-h_s^2/aL}\approx \frac{e^{2aL}}{2\pi aL}\int _{-\infty}^{\bar h-1}dh_se^{-h_s^2/aL}\\
&\quad=\sqrt{\frac{aL}{4}}\left[{\rm erf}\left(\frac{\bar h-1}{\sqrt{aL}}\right)+1\right]
\end{split}
\end{equation}

Computing the energy $\bar{H}_{in}$ of a fully solvated configuration
(belonging to the \textit{in} state) requires accounting for direct
interactions between solvent and solute, as well as counting broken
solvent-solvent bonds and the thermodynamic consequences of removing
a solvent cell to accommodate the solute. The four solvent-solute bonds
contribute an energy $-4 \epsilon_1$. Loss of 4 solvent-solvent
interactions is partially offset by the reversible work 
$\mu = -2 \epsilon_0$ of returning one solvent cell to the bath,
yielding a total energy
\begin{equation}
\label{eq:hamilsolvated}
\bar{H}_{in}= {\bar H_0} - 4(\epsilon_1-\epsilon_0/2) \,.
\end{equation}
The corresponding partition function can therefore be written as
\begin{equation}
Z_{in}(\bar{h}) =\mathop{\sum\dots\sum}_{h_1=0\,,h_L=0\,,h_s >\bar{h}}e^{-\beta \bar{H}_0}e^{\beta 4(\epsilon_1 - \epsilon_0/2)}
\end{equation}
Again, using Eq.~\ref{eq:ZM} in the limit $L>>1$, and rewriting the sums as integrals, we obtain
\begin{equation}
\label{eq:Zhs4}
\begin{split}
Z_{in}(\bar h)&\approx e^{\beta 4(\epsilon_1 - \epsilon_0/2)}\frac{e^{2aL}}{2\pi aL}\int ^{(L-1)/2}_{\bar h}dh_se^{-h_s^2/aL}\approx e^{\beta 4(\epsilon_1 - \epsilon_0/2)}\frac{e^{2aL}}{2\pi aL}\int ^{\infty}_{\bar h}dh_se^{-h_s^2/aL}\\
&\quad= e^{\beta 4(\epsilon_1 - \epsilon_0/2)}\sqrt{\frac{aL}{4}}{\rm erfc}\left(\frac{\bar h}{\sqrt{aL}}\right)
\end{split}
\end{equation}

Statistics of the pinned state are complicated by variation in the
solvent-solute interaction energy. Within the RSOS model, allowed
heights of columns $s-1$ and $s+1$ are $\bar{h}$, $\bar{h}-1$, and
$\bar{h}-2$. Only in the first case does the solute interact with an
adjacent liquid column. Since it is not necessary to break solvent-solvent bonds or
remove solvent cells in this case, the corresponding energy function
can be written as
\begin{equation}
\label{eq:barehamllpinned}
\bar{H}_{pinned} = \bar{H}_0 + \Delta H(h_{s-1},h_{s+1};\bar{h}),
\end{equation}
where
\begin{equation}
\Delta H(h_{s-1},h_{s+1};\bar{h}) = -\epsilon_1 (1 + \delta_{h_{s-1},\bar{h}}
+ \delta_{h_{s+1},\bar{h}}).
\end{equation}
Summation over the column heights $h_2, h_3, \ldots, h_{s-2}$
and $h_{s+2}, h_{s+3}, \ldots, h_{L-1}$ may now be performed using Eq.~\ref{eq:ZM}:
\begin{equation}
\begin{split}
&Z_{pinned} = \sum_{h_{s-1}=\bar{h}-2}^{\bar{h}} \sum_{h_{s+1}=\bar{h}-2}^{\bar{h}}
Z^{(1,s-1)}(0,h_{s-1}) \exp[-\beta\Delta H(h_{s-1},h_{s+1};\bar{h}) ]\times\\
&\quad\times
\exp\left[{ -\beta\epsilon_0 \over 2} \left(|h_{s-1}-(\bar{h}-1)|
+ |h_{s-1}-(\bar{h}-1)| \right)\right]Z^{(s+1,L)}(h_{s+1},0)\,.
\end{split}
\end{equation}
The constraint $h_s = \bar{h}-1$ defining the pinned state,
together with the locality of inter-column interactions, decouples
fluctuations to the left and right of the solute.  As a result,
summations over $h_{s-1}$ and $h_{s+1}$ can be carried out
independently; with the choice $s=(L+1)/2$ they yield identical
contributions in the $L\gg 1$ limit. We then obtain

\begin{figure}[tbp]
   \subfigure[$\,\,\epsilon_1=5\epsilon_0/12$]{
   \label{fig:Ep15}
   \includegraphics[scale=0.25,angle=0]{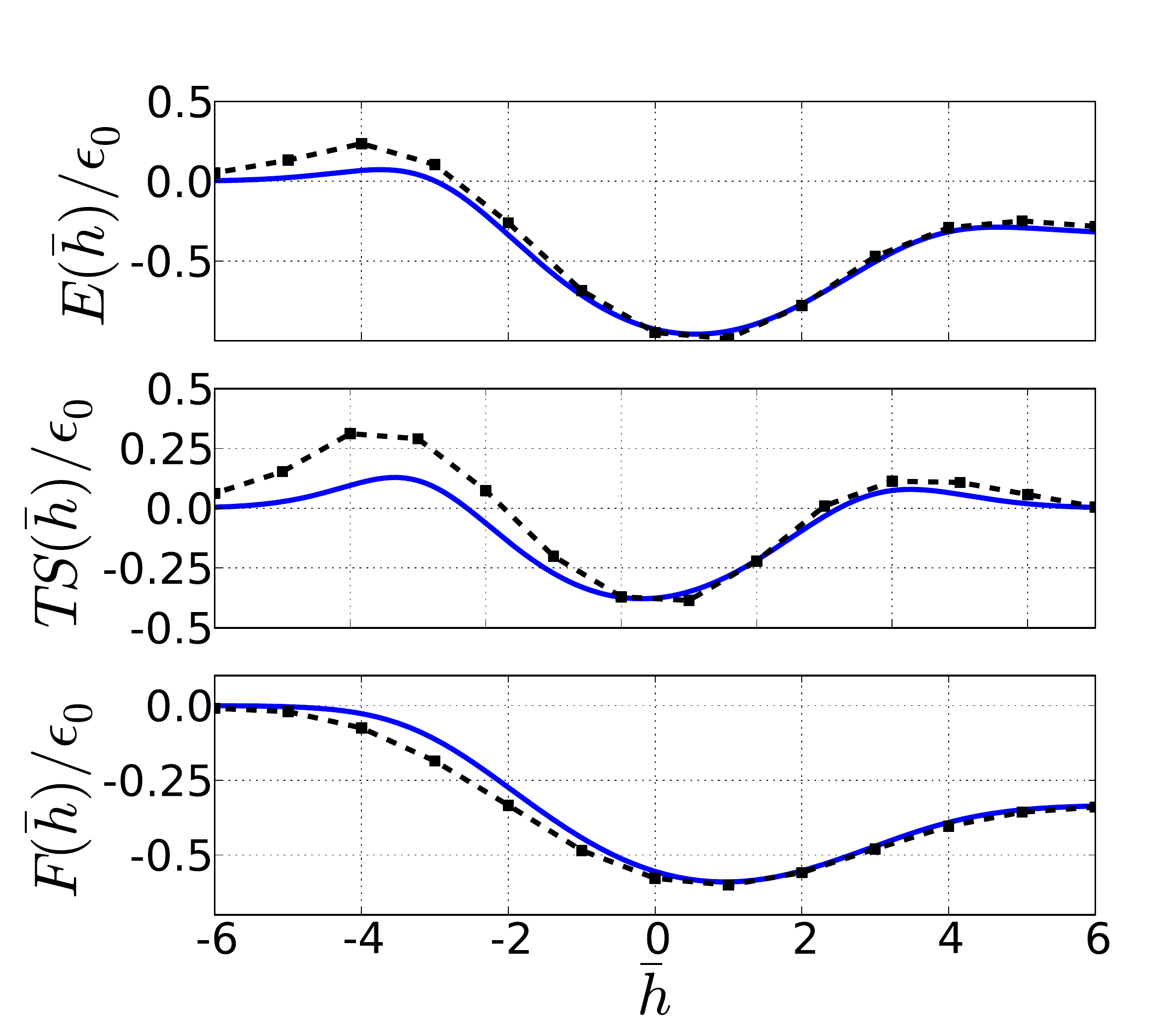}
   }
     \subfigure[$\, \epsilon_1=\epsilon_0/2$]{
   \label{fig:Ep18}
   \includegraphics[scale=0.25,angle=0]{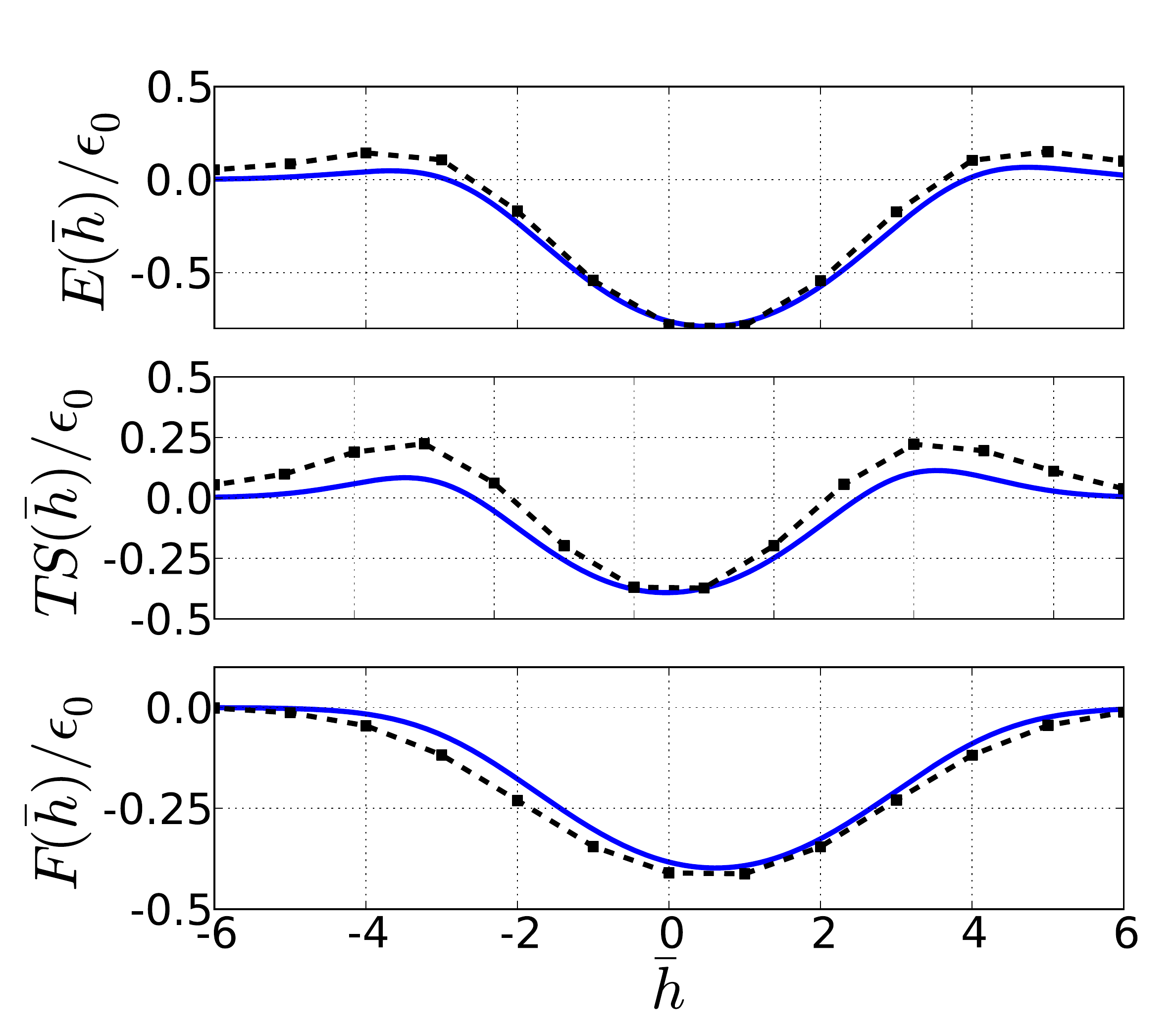}
   }
   \subfigure[$\, \epsilon_1=7\epsilon_0/12$]{
   \label{fig:Ep21}
   \includegraphics[scale=0.25,angle=0]{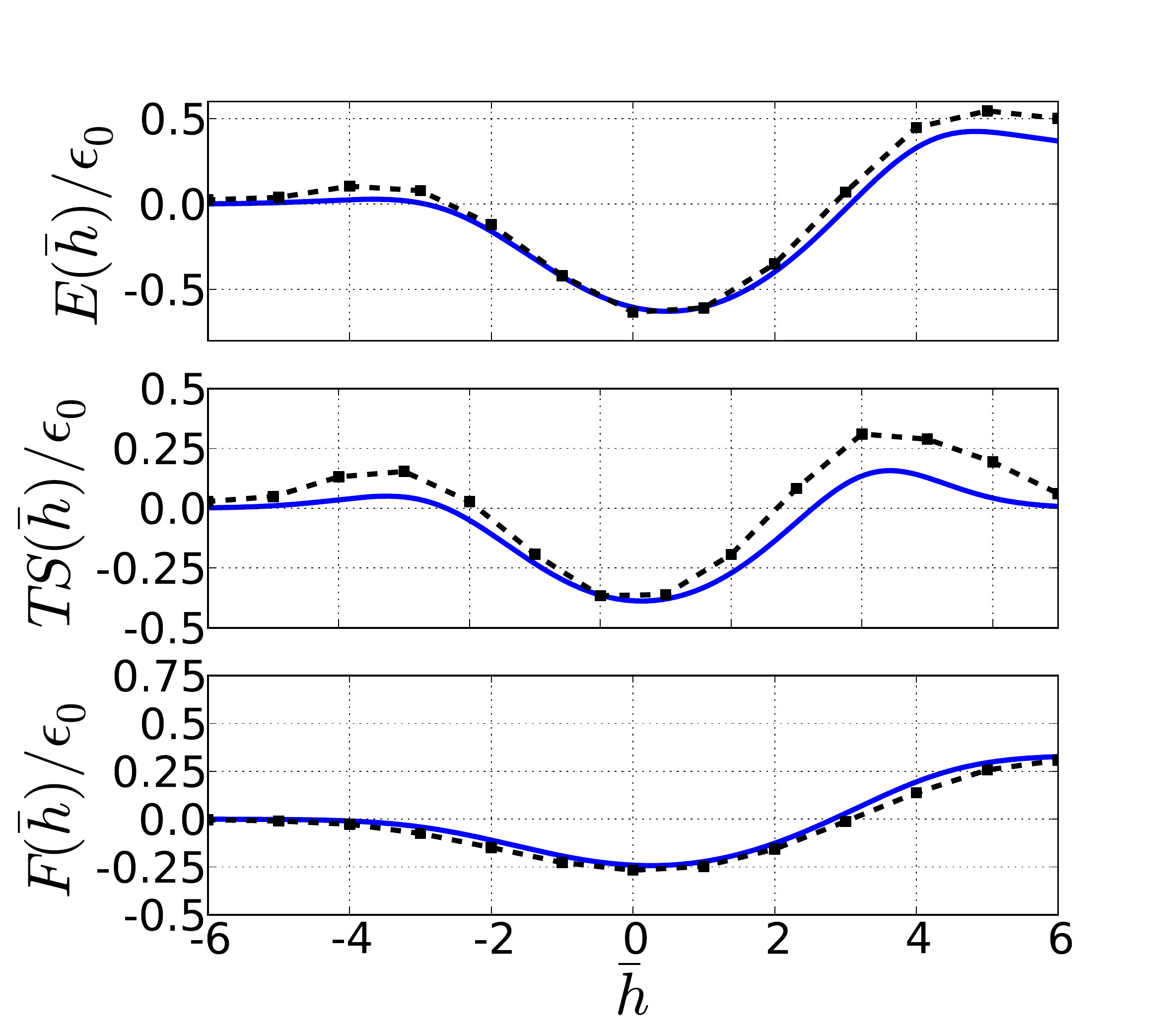}
   }
   \subfigure[$\, \epsilon_1=8\epsilon_0/12$]{
   \label{fig:Ep24}
   \includegraphics[scale=0.25,angle=0]{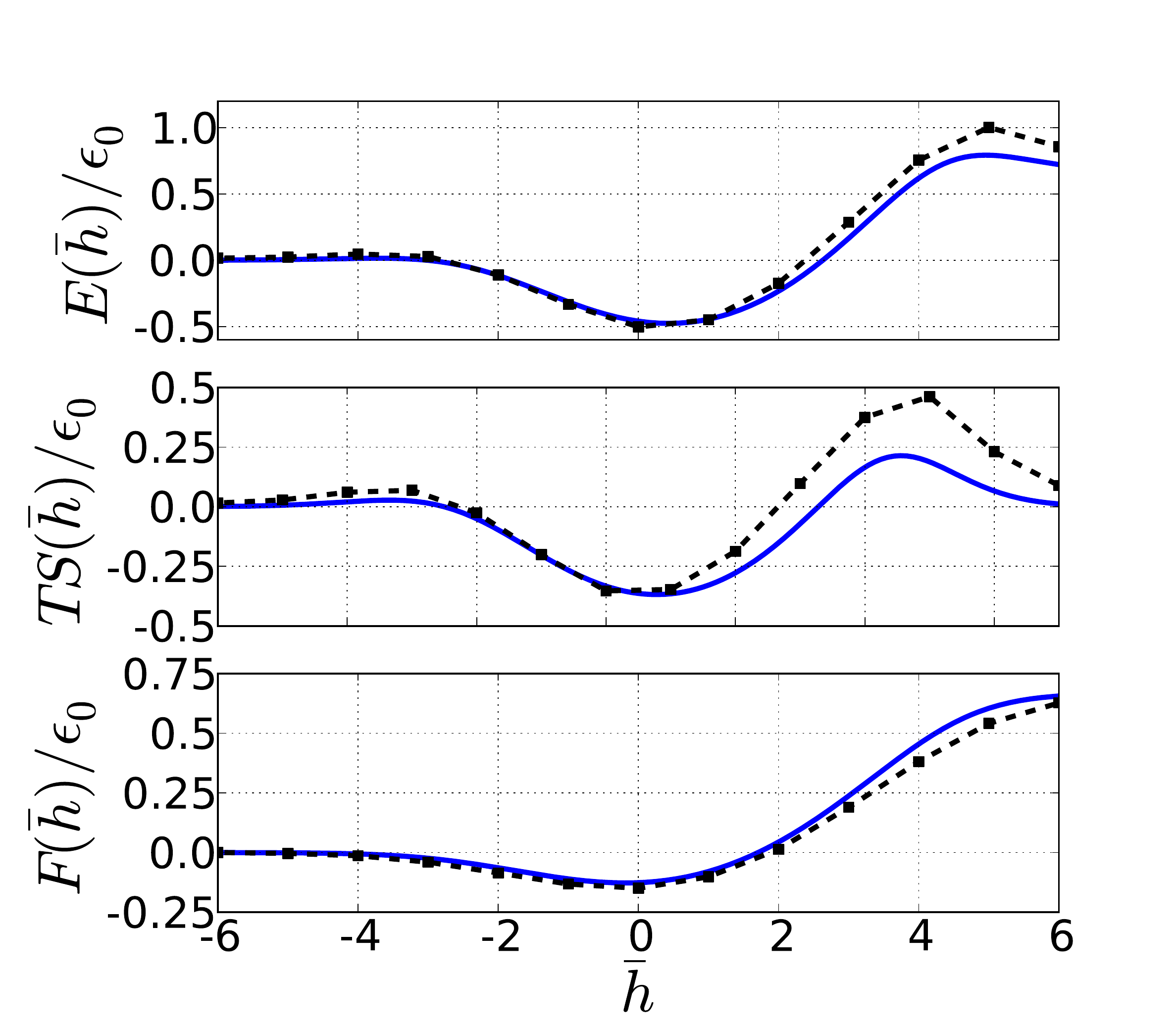}
   }
  \caption{Profiles of the  energy, entropy, and the free energy as a function of $\bar h$ for various values of $\epsilon_1$ at $T=\epsilon_0/3.6k_B$. The blue lines were calculated from Eq.~\ref{eq:Zhs}, while the dashed black lines with squares were obtained from simulations.}
\label{fig:1Dprofiles}
\end{figure}

\begin{equation}
\label{eq:Zhs5}
Z_{pinned}(\bar h)\approx b^2\frac{e^{2aL}}{2\pi aL}e^{-(\bar h-1)^2/aL}\,,
\end{equation} 
where 
\begin{equation}
b\equiv e^{-2a}\left[e^{\beta(3\epsilon_1/2-\epsilon_0/2)}e^{(-2\bar h+1)/(2aL)}+e^{\beta(\epsilon_1/2)}+e^{\beta((\epsilon_1-\epsilon_0)/2)}e^{(2\bar h-3)/(2aL)}\right]\,,
\end{equation}

Using Eqs.~\ref{eq:Zhs},\ref{eq:Zhs3},\ref{eq:Zhs4},\ref{eq:Zhs5}, we
can readily obtain profiles of free energy $F(\bar{h}) = -\beta^{-1}
\ln Z(\bar{h})$, energy $E(\bar{h}) = -(\partial \ln Z(\bar{h}) /
\partial \beta)_{L}$, and entropy $S(\bar{h}) = (E-F)/T$. These
analytical results are plotted in Fig.~\ref{fig:1Dprofiles} for a
lattice of size $L=30$, at temperature $T = \epsilon_0 / 3.6 k_{\rm
  B}$, and for various values of $\epsilon_1/\epsilon_0$. They agree
very well with simulation results.  
In particular, the depth of minima at $\bar{h}=0$ are accurately
predicted. In Fig.~\ref{fig:HalfEp1} we focus on these adsorption
properties, $\Delta g_{ads}\equiv g(0)-g(-(L-1)/2)$ for thermodynamic property $g$, for the specific case
$\epsilon_1=\epsilon_0/2$~\footnote{Recall that the bulk values of
  $F,E,S$ are equal in the liquid and vapor states when
  $\epsilon_1=\epsilon_0/2$. Hence, the results would not have changed
  if the bulk vapor was chosen as the reference state.}.  Theoretical
predictions are plotted as functions of temperature at $L=30$ alongside results
of Monte Carlo simulations. 

Agreement is very good for
temperatures in the 
range {$\epsilon_0/6k_B<T<\epsilon_0/3k_B$ for $L=30$}. 

The failure of our theory at high temperature stems from neglect of
contributions to $Z$ from terms of order $O(a^2)$ and higher.  At high
temperature ruggedness of the interface can also render the RSOS model
a poor proxy for the SOS model. At higher temperatures still, neglect of bulk inhomogeneities in the SOS model is itself not justified. 
{At low temperatures} a different prerequisite for the
validity of Eq.~\ref{eq:ZM} is violated, namely, $aL$ is no longer a
large number. {{For the RSOS fluid with $L=30$ columns, $aL\approx 1$ when $T\approx0.147\epsilon_0/k_B$. For temperatures around or lower than this value, height fluctuations in the RSOS fluid are no longer Gaussian to a good approximation}.  

Success of the SOS approximation 
in the temperature range $\epsilon_0/6k_B<T<\epsilon_0/3k_B$
reveals clearly the source of entropy
variations observed in our lattice gas simulations. Absent density
fluctuations in the liquid or vapor phase (which are neglected in the 
SOS description), fluctuations in interfacial shape provide the
sole source of entropy in these schematic models. The minimum 
in $S(\bar{h})$ near $\bar{h}=0$ thus directly indicates suppression
of capillary fluctuations when the solute resides at the liquid-vapor
interface. Similarly, local maxima of $S(\bar{h})$ just above and below
$\bar{h}=0$ signify enhancement of surface undulations when the solute
pulls the interface a few lattice spacings away from its natural
equilibrium position.
 \begin{figure}[tbp]
  \begin{center}
   \includegraphics[scale=0.35,angle=0]{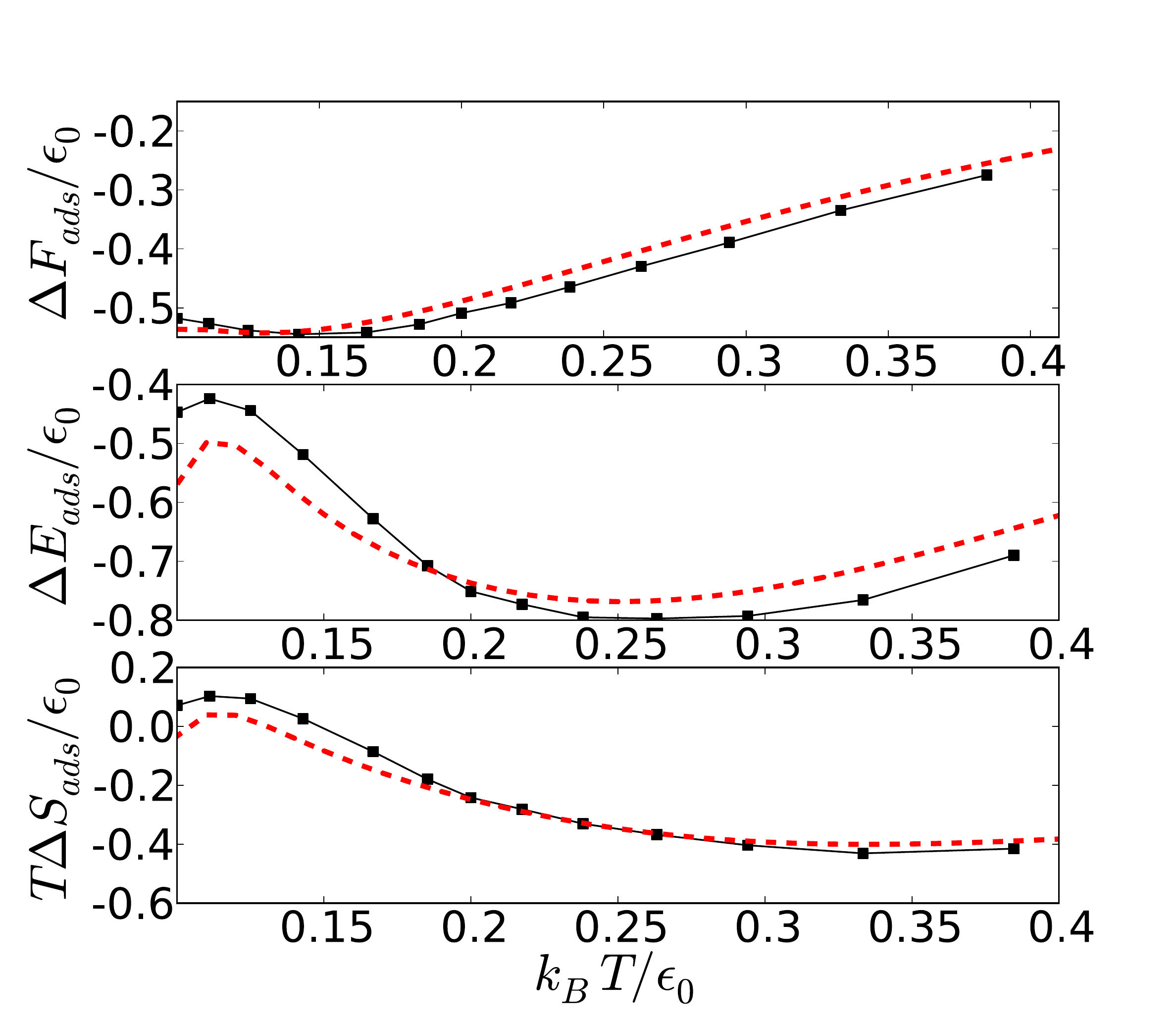}
      \caption{ Free energy, energy, entropy of adsorption as a function of $T$ for $\epsilon_1=\epsilon_0/2$, with $L=30$. The red line is calculated from theory, while the black line with circles is obtained from simulations.}
      \label{fig:HalfEp1}
      \end{center}
\end{figure}

Interfacial pinning is strongest for a solute with coupling constant
$\epsilon_1 \approx \epsilon_0/2$. In simulations this conclusion is
suggested by the depth of entropy minima at $\bar{h}=0$, which are
most pronounced at intermediate coupling. It can be drawn more
directly in our RSOS theory from the statistical weight
$P_{pinned}(\bar h)=Z_{pinned}(\bar h)/Z(\bar h)$ of the pinned
subensemble. For a given value of $\epsilon_1$, $P_{pinned}(\bar h)$
is always greatest at $\bar{h}=0$, where pinning does not require
deforming the interface. (See Fig.~\ref{fig:entdecompose1}) $P_{pinned}(0)$ should
clearly increase with $\epsilon_1$ in the strongly solvophobic regime,
where weak attraction to solvent is insufficient to offset the
entropic price of suppressing capillary-like waves. It is also clear
that $P_{pinned}(0)$ should decrease with $\epsilon_1$ in the strongly
solvophilic regime, where extremely favorable solvation encourages the
fluctuating interface to deform above a solute at $\bar{h}=0$ (i.e.,
to remain in the \textit{in} state).  Fig~\ref{fig:Ep2I} bears out the resulting
expectation that pinning is most effective at intermediate
$\epsilon_1$, very close to $\epsilon_1 = \epsilon_0/2$.

Subensemble weights $P_{\alpha}(\bar{h})\equiv
Z_\alpha(\bar{h})/Z(\bar{h})$, where $\alpha = \{in, out, pinned \}$
, also shed light on the nature of entropy maxima at 
values of $\bar{h}$ a few lattice spacings above and below the GDS. 
In the RSOS model the total entropy can be decomposed by
subensemble:
\begin{equation}
\label{eq:Entdecompose}
S(\bar h)=\sum_{\alpha} S_\alpha(\bar h) P_\alpha(\bar h)-k_B
\sum_\alpha P_\alpha(\bar h)\ln P_\alpha(\bar h)\,,
\end{equation}
where $S_{\alpha} = (E_{\alpha}-F_{\alpha})/T$ characterizes
fluctuations within each subensemble. The final term in
Eq.~\ref{eq:Entdecompose} is an entropy of mixing associated with
fluctuations between \textit{in}, \textit{out}, and \textit{pinned}
states.  Plots of $P_{\alpha}(\bar{h})$ in
Fig.~\ref{fig:entdecompose1} show that \textit{in} and \textit{pinned}
states become about equally probable near $\bar{h}=-2$, contributing a
mixing entropy of roughly $k_{\rm B}\ln 2$. This gain is partially
offset by the intrinsically low entropy of the \textit{pinned}
state, and by decreasing $S_{in}$ as the requirement $h_s > \bar{h}$
becomes a nonnegligible constraint on interfacial fluctuations in the
solvated state. 
The residual increase in entropy is a signature
of spontaneous switching from the pinned state to an unpinned state,
and vice versa. Analogous switching between pinned and unsolvated
states is responsible for the entropy maximum at $\bar{h}= + 4$.

\begin{figure}[h!]
   \subfigure[]{
   \label{fig:entdecompose1}
   \includegraphics[scale=0.25,angle=0]{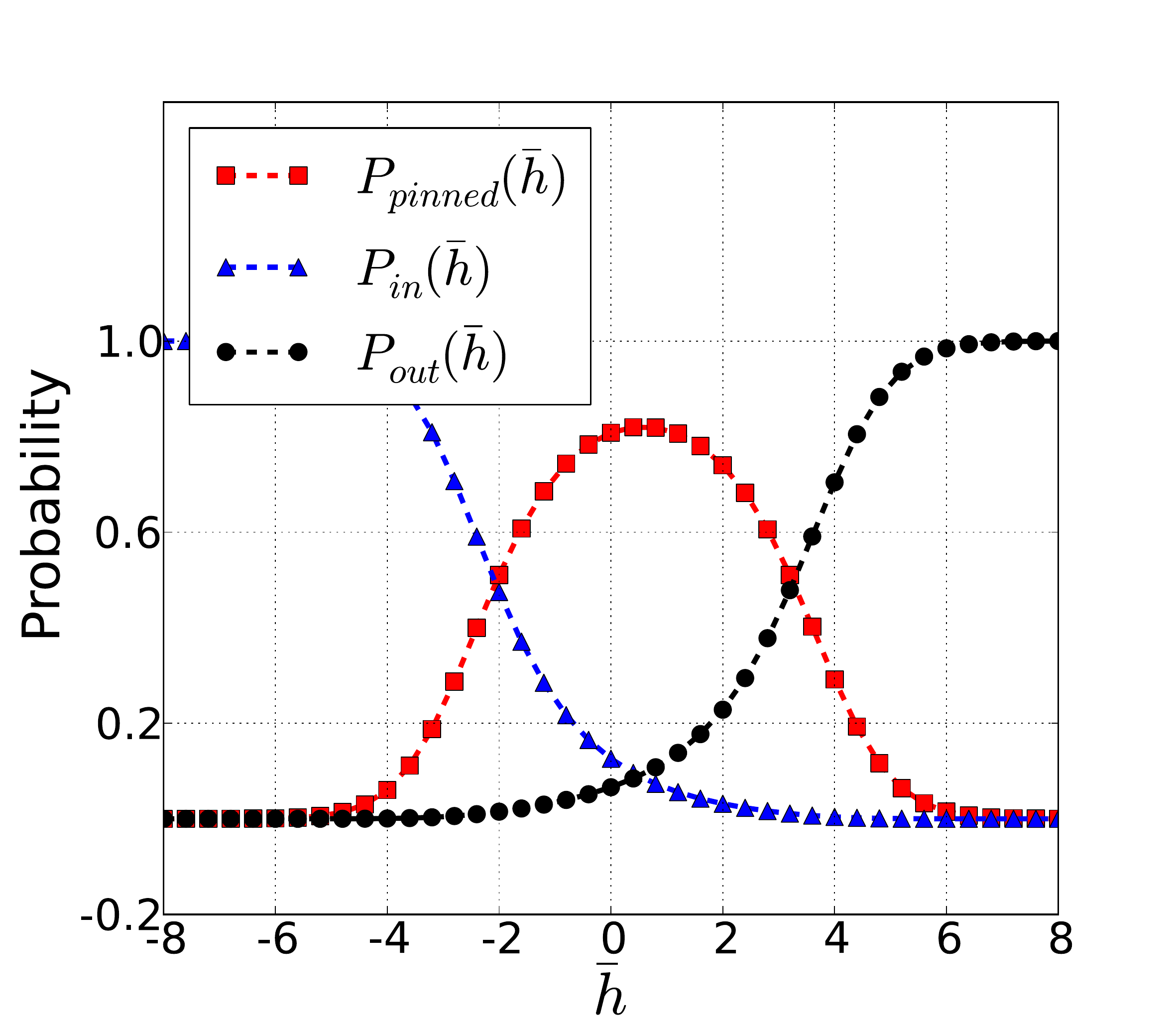}
   }
   \subfigure[]{
   \label{fig:Ep2I}
   \includegraphics[scale=0.25,angle=0]{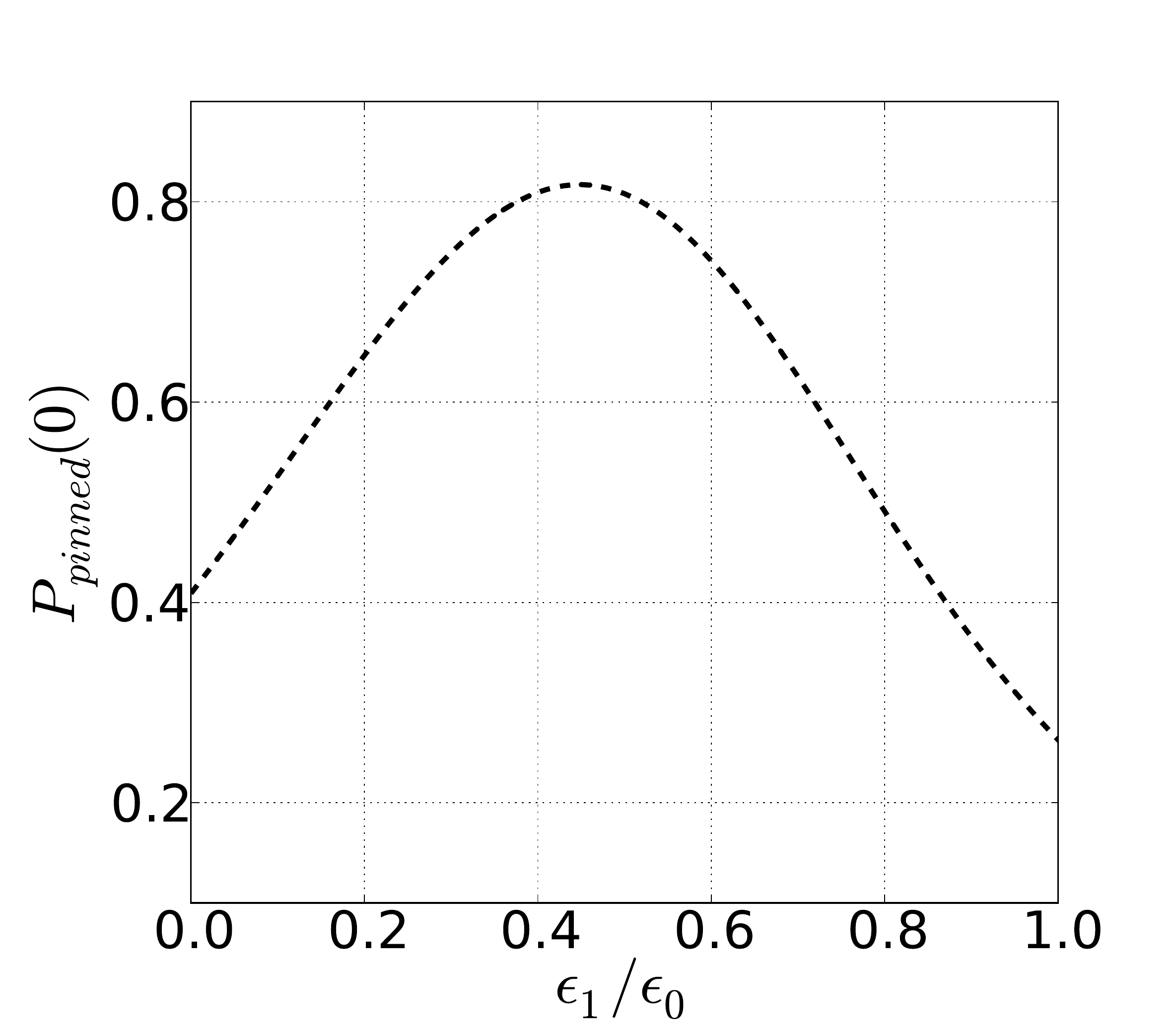}
   }
   
  \caption{(a) Probabilities of the \textit{pinned}, \textit{in}, and \textit{out} states as a function of $\bar h$ for $T=\epsilon_0/3.6k_B$, $\epsilon_1/\epsilon_0=1/2$ and (b) probability of the \textit{pinned} state at $\bar h=0$ as a function of $\epsilon_1$ for $T=\epsilon_0/3.6k_B$. When the solute is near the GDS at $\bar h=0$, most of the configurations are in the \textit{pinned} state.}
\label{fig:Entdecompose}
\end{figure}

Energy can also be decomposed by subensemble:
\begin{equation}
\label{eq:Engdecompose}
E(\bar h)=\sum_{\alpha} E_\alpha(\bar h)P_\alpha(\bar h)\,,
\end{equation}
where $E_\alpha(\bar h)$ is the average energy in subensemble $\alpha$
when the solute is held at
$\bar{h}$. Eqs~\ref{eq:barehamil},~\ref{eq:hamilsolvated},~\ref{eq:barehamllpinned},
allow rough estimates of $E_\alpha(\bar h)$ and thus simple
predictions for energies of adsorption. For $\bar{h}\gg 0$ the
unsolvated state dominates ($P_{out}= 1$), giving $E(\infty) \approx
E_{out} \approx E_0$, where $E_0$ is the average energy of the
solute-free system. For $\bar{h}\ll 0$ the solvated state dominates
($P_{in}= 1$), giving $E(-\infty) \approx E_{in} \approx E_0 -
4(\epsilon_1 - \epsilon_0/2)$ from Eq.~\ref{eq:hamilsolvated}. To the
extent that pinning dominates at $\bar{h} = 0$, we have $E(0) \approx
E_{pinned}(0) \approx E_0 +\Delta H(h_{s-1},h_{s+1};\bar{h})$.  At 
temperatures low enough to safely set
$h_{s-1}=h_{s+1}=\bar{h}$ for the pinned state, we have $E(0) \approx
E_0 -3\epsilon_1-\epsilon_0$. Interfacial adsorption of the solute
from the liquid phase therefore produces an energy change $E(0) - E(-\infty) \approx \epsilon_1 - \epsilon_0$,
accounting roughly for replacing a solvent-solute bond with a
solvent-solvent bond. For a modestly solvophilic solute ($\epsilon_0/2
< \epsilon_1 < 1)$, adsorption is energetically favorable, and the
energetic driving force strengthens with decreasing $\epsilon_1$.
This force is stronger still in the solvophobic regime ($\epsilon_1 <
\epsilon_0/2$), but these solutes reside primarily in the vapor
phase. The strongest relevant energy of adsorption from the liquid
phase is thus obtained for $\epsilon_1 \approx \epsilon_0/2$, for
which $\Delta E_{ads} \approx -\epsilon_0/2$. 

These crude predictions are consistent at low temperatures (below $T=0.15\,\epsilon_0/k_B$ for $L=30$) with results
from our simulations and detailed theoretical analysis (see
Fig.~\ref{fig:HalfEp1}). At temperatures above $T=0.15\,\epsilon_0/k_B$ for a lattice of size $L=30$, interfacial fluctuations modify $\Delta E_{ads}$
significantly. Reduced coordination at a rough interface enhances the
energetic driving force for adsorption, effectively increasing the
number of solvent-solute and solvent-solvent bonds exchanged when the
solute moves to the surface (see Fig.~\ref{fig:HalfEp1}).  At the same
time, however, the opposing entropic forces discussed above become
more pronounced as temperature increases. It is not obvious from our
simple estimates how this competition plays out, though the sign and magnitude of $\Delta S_{ads}$ ensures that net surface propensity decreases smoothly with increasing
temperature (see Fig.~\ref{fig:HalfEp1}).
Regardless of temperature, adsorption is always strongest when
$\epsilon_1=\epsilon_0/2$ (see Fig.~\ref{fig:maxadsorption}).

 \begin{figure}[tbp]
   \begin{center}
   \includegraphics[scale=0.35,angle=0]{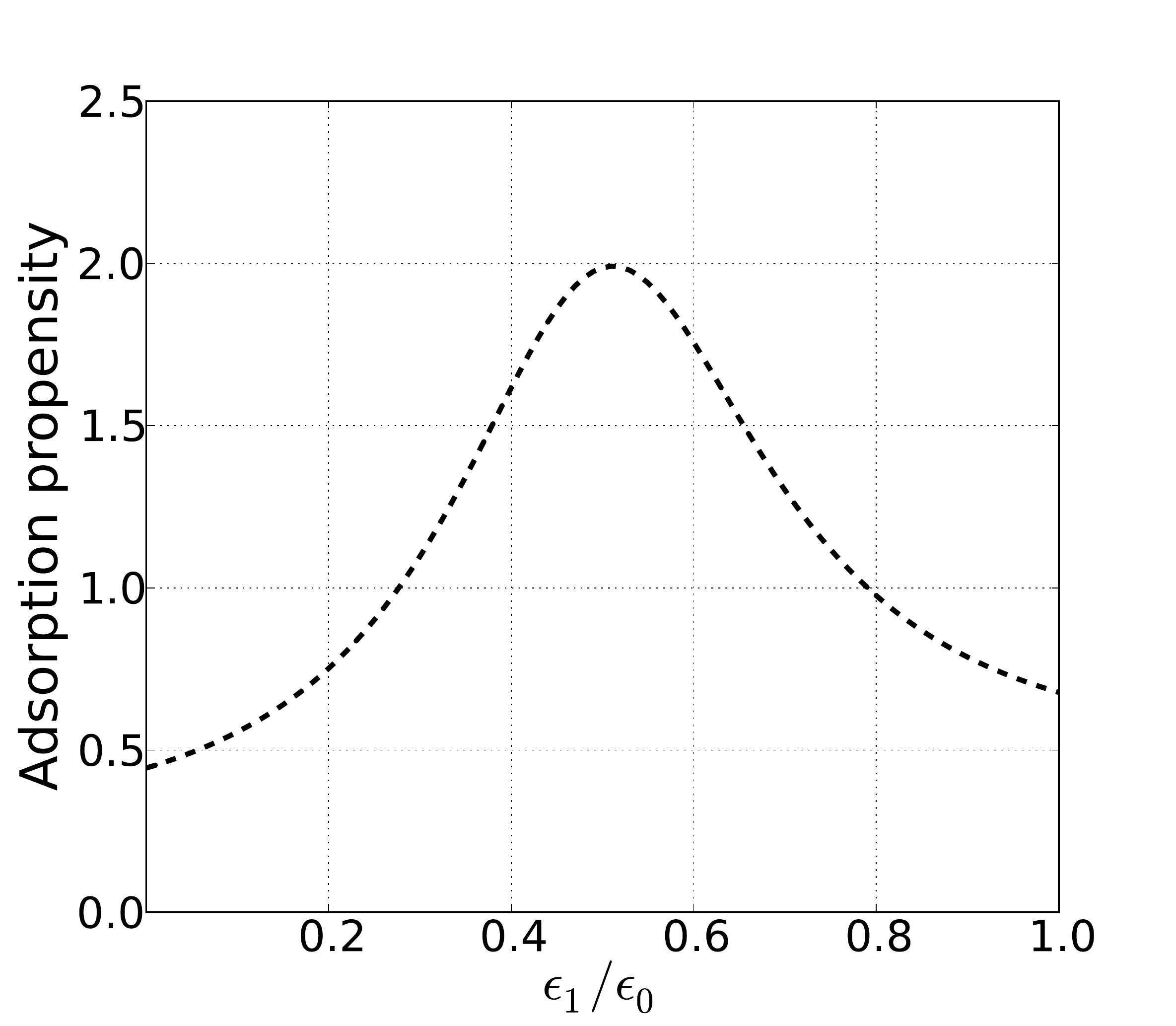}
  \caption{Adsorption propensity defined as $\frac{Z_{}(0)}{Z({-(L-1)}/{2})+Z_{}({(L-1)}/{2})}$ as a function of $\epsilon_1$ for $T=\epsilon_0/3.6k_B$. These estimates were calculated using Eq.~\ref{eq:Zhs}.}
\label{fig:maxadsorption}
\end{center}
\end{figure}

\section{Conclusions}
\label{sec:Conclusion}

Among models that resolve microscopic fluctuations of the liquid-vapor
interface, the two-dimensional lattice gas we have
studied may be the simplest from which mechanisms of solute surface
propensity might be learned. Any insight it offers into the behavior
of real molecular systems is no doubt schematic. But the fact that it
captures thermodynamic trends which elude celebrated theories of
solvation encourages a detailed view of its conclusions in the context
of more realistic molecular models.

Any such comparison requires assigning interaction parameters
$\epsilon_0$ and $\epsilon_1$ to systems with much more complicated
energetics. We do so through the local approximation of
Eq.~\ref{eq:Ulocal}, which for simulations accurately describes the potential energy
of a non-polarizable ion in water at a level of detail akin to that of a lattice gas.
For a three-dimensional lattice gas model of bulk solvent, the total
energy could be calculated by attributing to each liquid cell half of
the interactions with its 6 neighboring liquid cells, so that
$\epsilon_{bulk} = -3\epsilon_0$ (neglecting solvent density
inhomogeneities). Similarly, since solvent cells at an ideally flat
liquid-vapor interface engage in one fewer interaction than in bulk,
we estimate $\epsilon_{interface} \approx -(5/2) \epsilon_0$.  From
the value of $\epsilon_{interface}-\epsilon_{bulk} = 6.52$ kJ/mol
computed from atomistic simulations of water~\cite{Otten:2012}, we thus infer a
coarse-grained solvent-solvent attraction strength $\epsilon_0 \approx
13.04$ kJ/mol.  Within the local approximation,
solvent cells adjacent to the solute would be assigned half of their
solvent-solvent interactions as well as a single solute bond. For a
solute well within bulk solution, we then have $\epsilon_{coord} =
-\epsilon_1 - (5/2) \epsilon_0$ in the absence of solvent density
inhomogeneities (i.e., within the SOS approximation).  From the value
$\epsilon_{coord}-\epsilon_{bulk} = -4.44$ kJ/mol determined for
a fractionally charged non-polarizable Iodine ion in water, I$^{-0.8}$, by simulation~\cite{Otten:2012} we finally obtain $\epsilon_1 = 10.96$
kJ/mol, or $\epsilon_1/\epsilon_0 \approx 0.84$.

From a lattice gas perspective, the simulated solute I$^{-0.8}$ (aq)
therefore falls within the hydrophilic regime $\epsilon_1/\epsilon_0 >
1/2$. Further reducing the magnitude of the solute's charge should
certainly weaken solute-solvent attractions. Decreasing $\epsilon_1$
by a small amount in this way should bring the solute closer to
conditions of maximum adsorption, $\epsilon_1 = \epsilon_0/2$, where
the scales of adsorption energy and entropy are both largest.  Based
on results we have presented, we thus expect that changing the ion's
charge from $-0.8 e$ to $-0.75 e$ would deepen the interfacial minima
of energy and entropy profiles.  Using the same molecular simulation
techniques described in Ref.~\cite{Otten:2012}, we find this
prediction to indeed hold true.

In drawing this comparison we have assumed that basic interfacial
solvation properties of the lattice gas are insensitive to
dimensionality $d$. Although scaling behavior of a lattice gas near
the critical point is famously sensitive to $d$, there is good reason
to expect that the essence of our conclusions regarding solvation will
be unchanged when moving from $d=2$ to $d=3$.  The liquid-vapor
interface of a three-dimsional lattice gas similarly supports
capillary wave-like fluctuations for temperatures in the range
$T_R<T<T_c$, where $T_R$
denotes the so-called roughening transition
temperature~\cite{Weeks:1973}. Furthermore, these surface undulations
are Gaussian distributed to a good approximation at long
wavelengths~\cite{Unpublished1,Weeks:1977}, just as in
Eq.~\ref{eq:ZM}.  Partition functions of the {\em in}, {\em out}, and
{\em pinned} states should therefore have similar forms in $d=3$.  In
qualitative terms, only the scale of height fluctuations governing
$Z_{in}$, $Z_{out}$, and $Z_{pinned}$ will change, growing with system
size only as $\sqrt{\ln L}$ rather than $\sqrt{L}$
~\cite{Safran:2003}. Preliminary simulation results for
the $d=3$ case bear out these expectations.

We conclude that lattice gas models of interfacial solvation are a
promising starting point for scrutinizing essential effects of surface
roughness and energetic heterogeneity on solute adsorption. For the
especially interesting case of charged solutes in highly polar
solvents, a more realistic treatment of electrostatics is clearly in
order. Given the success of lattice models for dielectric response
~\cite{Song:1998}, this improvement appears
quite feasible. An ability to treat solutes of varying size and
geometry would also greatly aid connections with experiment. These
developments are underway.

\appendix
\section{Appendix}
Here we present arguments underlying the assertion that a solute with
coupling constant $\epsilon_1=\epsilon_0/2$ is equally likely to
reside in either bulk phase.
Consider a pair of configurations $\{ n_i, n_i^s \}$ and $\{ n_i',
n_i^s \}$ related by a global transformation that swaps liquid and
vapor phases. Specifically, the configuration $\{ n_i', n_i^s \}$ is
obtained from $\{ n_i, n_i^s \}$ by changing the solvent occupation
state of all lattice cells not occupied by the solute, $n_i' =
(1-n_i^s)(1-n_i)$. Repeating this transformation, $n_i =
(1-n_i^s)(1-n_i')$, we recover the original configuration $\{ n_i,
n_i^s \}$. In simple terms, if the solute resides in the liquid phase
in $\{ n_i, n_i^s \}$, then it resides in the vapor phase in $\{ n_i',
n_i^s \}$, and vice versa.

With an accompanying transformation of the solute-solvent coupling
$\epsilon_1$, the energies of configurations $\{ n_i, n_i^s \}$ and
$\{ n_i^\prime, n_i^s \}$ can be directly related. Defining $E(\{ n_i, n_i^s
\}; \{\epsilon_0,\epsilon_1\})$ as the energy of configuration $\{ n_i, n_i^s \}$
with solvent-solvent coupling $\epsilon_0$, and solvent-solute coupling $\epsilon_1$, we have%

\begin{equation}
\label{eq:latticegasappendix2}
\begin{split}
E(\{n_i^\prime,n^s_i\};\{\epsilon_0,\epsilon_1\})&=-{\epsilon_0}\sum_{i,j\in nn} n_i^\prime n_j^\prime-\mu\sum_{i}n_i^\prime-\epsilon_1\sum_{i,j\in nn} n^s_in_j^\prime\quad\\&=-{\epsilon_0}\sum_{i,j\in nn} n_in_j-\mu\sum_{i}n_i-(\epsilon_0-\epsilon_1)\sum_{i,j\in nn} n^s_in_j+4(\epsilon_1-\epsilon_0/2)\\
&=E(\{n_i,n^s_i\};\{\epsilon_0,\epsilon_0-\epsilon_1\})+4(\epsilon_1-\epsilon_0/2). 
\end{split}
\end{equation}
When $\epsilon_1=\epsilon_0/2$, the energy of the two configurations are the same. Thus for every configuration in which the solute is in the liquid phase, there exists a configuration with the same statistical weight in which the solute is in the vapor phase and vice versa. Hence solutes with $\epsilon_1=\epsilon_0/2$ exhibit equal preference for the liquid and vapor phases.

\section*{Acknowledgements}
Kelsey Schuster performed preliminary simulations on the lattice gas model. We gratefully acknowledge useful discussions with Yan Levin and David Limmer. This project was supported by the US Department of Energy, Office of Basic Energy Sciences, through the Chemical Sciences Division (CSD) of the Lawrence Berkeley National Laboratory (LBNL), under Contract DE-AC02-05CH11231.

\providecommand*{\mcitethebibliography}{\thebibliography}
\csname @ifundefined\endcsname{endmcitethebibliography}
{\let\endmcitethebibliography\endthebibliography}{}


\begin{mcitethebibliography}{26}
\providecommand*{\natexlab}[1]{#1}
\providecommand*{\mciteSetBstSublistMode}[1]{}
\providecommand*{\mciteSetBstMaxWidthForm}[2]{}
\providecommand*{\mciteBstWouldAddEndPuncttrue}
  {\def\EndOfBibitem{\unskip.}}
\providecommand*{\mciteBstWouldAddEndPunctfalse}
  {\let\EndOfBibitem\relax}
\providecommand*{\mciteSetBstMidEndSepPunct}[3]{}
\providecommand*{\mciteSetBstSublistLabelBeginEnd}[3]{}
\providecommand*{\EndOfBibitem}{}
\mciteSetBstSublistMode{f}
\mciteSetBstMaxWidthForm{subitem}
{(\emph{\alph{mcitesubitemcount}})}
\mciteSetBstSublistLabelBeginEnd{\mcitemaxwidthsubitemform\space}
{\relax}{\relax}

\bibitem[Jungwirth and Tobias(2006)]{Jungwirth:2006}
P.~Jungwirth and D.~J. Tobias, \emph{Chem. Rev.}, 2006, \textbf{106},
  1259--1281\relax
\mciteBstWouldAddEndPuncttrue
\mciteSetBstMidEndSepPunct{\mcitedefaultmidpunct}
{\mcitedefaultendpunct}{\mcitedefaultseppunct}\relax
\EndOfBibitem
\bibitem[Jungwirth and Tobias(2002)]{Jungwirth:2002}
P.~Jungwirth and D.~Tobias, \emph{The Journal of Physical Chemistry B}, 2002,
  \textbf{106}, 6361--6373\relax
\mciteBstWouldAddEndPuncttrue
\mciteSetBstMidEndSepPunct{\mcitedefaultmidpunct}
{\mcitedefaultendpunct}{\mcitedefaultseppunct}\relax
\EndOfBibitem
\bibitem[Petersen and Saykally(2004)]{Petersen:2004}
P.~Petersen and R.~Saykally, \emph{Chemical Physics Letters}, 2004,
  \textbf{397}, 51--55\relax
\mciteBstWouldAddEndPuncttrue
\mciteSetBstMidEndSepPunct{\mcitedefaultmidpunct}
{\mcitedefaultendpunct}{\mcitedefaultseppunct}\relax
\EndOfBibitem
\bibitem[Noah-Vanhoucke and Geissler(2009)]{NoahVanhoucke:2009}
J.~Noah-Vanhoucke and P.~Geissler, \emph{Proceedings of the National Academy of
  Sciences}, 2009, \textbf{106}, 15125--15130\relax
\mciteBstWouldAddEndPuncttrue
\mciteSetBstMidEndSepPunct{\mcitedefaultmidpunct}
{\mcitedefaultendpunct}{\mcitedefaultseppunct}\relax
\EndOfBibitem
\bibitem[Otten \emph{et~al.}(2012)Otten, Shaffer, Geissler, and
  Saykally]{Otten:2012}
D.~Otten, P.~Shaffer, P.~Geissler and R.~Saykally, \emph{Proceedings of the
  National Academy of Sciences}, 2012, \textbf{109}, 701--705\relax
\mciteBstWouldAddEndPuncttrue
\mciteSetBstMidEndSepPunct{\mcitedefaultmidpunct}
{\mcitedefaultendpunct}{\mcitedefaultseppunct}\relax
\EndOfBibitem
\bibitem[Onsager and Samaras(1934)]{Onsager:1934}
L.~Onsager and N.~Samaras, \emph{The Journal of Chemical Physics}, 1934,
  \textbf{2}, 528\relax
\mciteBstWouldAddEndPuncttrue
\mciteSetBstMidEndSepPunct{\mcitedefaultmidpunct}
{\mcitedefaultendpunct}{\mcitedefaultseppunct}\relax
\EndOfBibitem
\bibitem[Archontis and Leontidis(2006)]{Archontis:2006}
G.~Archontis and E.~Leontidis, \emph{Chemical Physics Letters}, 2006,
  \textbf{420}, 199--203\relax
\mciteBstWouldAddEndPuncttrue
\mciteSetBstMidEndSepPunct{\mcitedefaultmidpunct}
{\mcitedefaultendpunct}{\mcitedefaultseppunct}\relax
\EndOfBibitem
\bibitem[Levin(2009)]{Levin:2009}
Y.~Levin, \emph{Physical Review Letters}, 2009, \textbf{102}, 147803\relax
\mciteBstWouldAddEndPuncttrue
\mciteSetBstMidEndSepPunct{\mcitedefaultmidpunct}
{\mcitedefaultendpunct}{\mcitedefaultseppunct}\relax
\EndOfBibitem
\bibitem[Levin \emph{et~al.}(2009)Levin, Santos, and Diehl]{Levin:2009b}
Y.~Levin, A.~P.~D. Santos and A.~Diehl, \emph{Physical Review Letters}, 2009,
  \textbf{103}, 257802\relax
\mciteBstWouldAddEndPuncttrue
\mciteSetBstMidEndSepPunct{\mcitedefaultmidpunct}
{\mcitedefaultendpunct}{\mcitedefaultseppunct}\relax
\EndOfBibitem
\bibitem[Markin and Volkov(2002)]{Markin:2002}
V.~Markin and A.~Volkov, \emph{The Journal of Physical Chemistry B}, 2002,
  \textbf{106}, 11810--11817\relax
\mciteBstWouldAddEndPuncttrue
\mciteSetBstMidEndSepPunct{\mcitedefaultmidpunct}
{\mcitedefaultendpunct}{\mcitedefaultseppunct}\relax
\EndOfBibitem
\bibitem[Huang and Chandler(2002)]{Huang:2002}
D.~M. Huang and D.~Chandler, \emph{The Journal of Physical Chemistry B}, 2002,
  \textbf{106}, 2047--2053\relax
\mciteBstWouldAddEndPuncttrue
\mciteSetBstMidEndSepPunct{\mcitedefaultmidpunct}
{\mcitedefaultendpunct}{\mcitedefaultseppunct}\relax
\EndOfBibitem
\bibitem[Caleman \emph{et~al.}(2011)Caleman, Hub, van Maaren, and van~der
  Spoel]{Caleman:2011}
C.~Caleman, J.~Hub, P.~van Maaren and D.~van~der Spoel, \emph{Proceedings of
  the National Academy of Sciences}, 2011, \textbf{108}, 6838\relax
\mciteBstWouldAddEndPuncttrue
\mciteSetBstMidEndSepPunct{\mcitedefaultmidpunct}
{\mcitedefaultendpunct}{\mcitedefaultseppunct}\relax
\EndOfBibitem
\bibitem[Iuchi \emph{et~al.}(2009)Iuchi, Chen, Paesani, and Voth]{Iuchi:2009}
S.~Iuchi, H.~Chen, F.~Paesani and G.~A. Voth, \emph{The Journal of Physical
  Chemistry B}, 2009, \textbf{113}, 4017--4030\relax
\mciteBstWouldAddEndPuncttrue
\mciteSetBstMidEndSepPunct{\mcitedefaultmidpunct}
{\mcitedefaultendpunct}{\mcitedefaultseppunct}\relax
\EndOfBibitem
\bibitem[Stuart and Berne(1996)]{Stuart:1996}
S.~Stuart and B.~Berne, \emph{The Journal of Physical Chemistry}, 1996,
  \textbf{100}, 11934--11943\relax
\mciteBstWouldAddEndPuncttrue
\mciteSetBstMidEndSepPunct{\mcitedefaultmidpunct}
{\mcitedefaultendpunct}{\mcitedefaultseppunct}\relax
\EndOfBibitem
\bibitem[Vaitheeswaran and Thirumalai(2006)]{Vaitheeswaran:2006}
S.~Vaitheeswaran and D.~Thirumalai, \emph{Journal of the American Chemical
  Society}, 2006, \textbf{128}, 13490--13496\relax
\mciteBstWouldAddEndPuncttrue
\mciteSetBstMidEndSepPunct{\mcitedefaultmidpunct}
{\mcitedefaultendpunct}{\mcitedefaultseppunct}\relax
\EndOfBibitem
\bibitem[Weeks(1977)]{Weeks:1977}
J.~Weeks, \emph{The Journal of Chemical Physics}, 1977, \textbf{67}, 3106\relax
\mciteBstWouldAddEndPuncttrue
\mciteSetBstMidEndSepPunct{\mcitedefaultmidpunct}
{\mcitedefaultendpunct}{\mcitedefaultseppunct}\relax
\EndOfBibitem
\bibitem[Chui and Weeks(1981)]{Chui:1981}
S.~Chui and J.~Weeks, \emph{Phys. Rev. B}, 1981, \textbf{23}, 2438\relax
\mciteBstWouldAddEndPuncttrue
\mciteSetBstMidEndSepPunct{\mcitedefaultmidpunct}
{\mcitedefaultendpunct}{\mcitedefaultseppunct}\relax
\EndOfBibitem
\bibitem[Abraham(1980)]{Abraham:1980}
D.~B. Abraham, \emph{Phys. Rev. Lett.}, 1980, \textbf{44}, 1165--1168\relax
\mciteBstWouldAddEndPuncttrue
\mciteSetBstMidEndSepPunct{\mcitedefaultmidpunct}
{\mcitedefaultendpunct}{\mcitedefaultseppunct}\relax
\EndOfBibitem
\bibitem[Fisher(1984)]{Fisher:1984}
M.~Fisher, \emph{Journal of Statistical Physics}, 1984, \textbf{34},
  667--729\relax
\mciteBstWouldAddEndPuncttrue
\mciteSetBstMidEndSepPunct{\mcitedefaultmidpunct}
{\mcitedefaultendpunct}{\mcitedefaultseppunct}\relax
\EndOfBibitem
\bibitem[Nelson \emph{et~al.}(2004)Nelson, Piran, and Weinberg]{Nelson:2004}
D.~Nelson, T.~Piran and S.~Weinberg, \emph{Statistical Mechanics of Membranes
  and Surfaces}, World Scientific Pub., 2004\relax
\mciteBstWouldAddEndPuncttrue
\mciteSetBstMidEndSepPunct{\mcitedefaultmidpunct}
{\mcitedefaultendpunct}{\mcitedefaultseppunct}\relax
\EndOfBibitem
\bibitem[Temperley(1952)]{Temperley:1952}
H.~N.~V. Temperley, \emph{Mathematical Proceedings of the Cambridge
  Philosophical Society}, 1952,  683--697\relax
\mciteBstWouldAddEndPuncttrue
\mciteSetBstMidEndSepPunct{\mcitedefaultmidpunct}
{\mcitedefaultendpunct}{\mcitedefaultseppunct}\relax
\EndOfBibitem
\bibitem[Safran(2003)]{Safran:2003}
S.~Safran, \emph{Statistical Thermodynamics Of Surfaces, Interfaces, And
  Membranes}, Westview Press, 2003\relax
\mciteBstWouldAddEndPuncttrue
\mciteSetBstMidEndSepPunct{\mcitedefaultmidpunct}
{\mcitedefaultendpunct}{\mcitedefaultseppunct}\relax
\EndOfBibitem
\bibitem[Kardar(2007)]{Kardar2007}
M.~Kardar, \emph{Statistical Physics of Fields}, Cambridge University Press,
  2007\relax
\mciteBstWouldAddEndPuncttrue
\mciteSetBstMidEndSepPunct{\mcitedefaultmidpunct}
{\mcitedefaultendpunct}{\mcitedefaultseppunct}\relax
\EndOfBibitem
\bibitem[Weeks \emph{et~al.}(1973)Weeks, Gilmer, and Leamy]{Weeks:1973}
J.~Weeks, G.~Gilmer and H.~Leamy, \emph{Physical Review Letters}, 1973,
  \textbf{31}, 549--551\relax
\mciteBstWouldAddEndPuncttrue
\mciteSetBstMidEndSepPunct{\mcitedefaultmidpunct}
{\mcitedefaultendpunct}{\mcitedefaultseppunct}\relax
\EndOfBibitem
\bibitem[Vaikuntanathan \emph{et~al.}(2012)Vaikuntanathan, Shaffer, and
  Geissler]{Unpublished1}
S.~Vaikuntanathan, P.~Shaffer and P.~L. Geissler, \emph{Unpublished},
  2012\relax
\mciteBstWouldAddEndPuncttrue
\mciteSetBstMidEndSepPunct{\mcitedefaultmidpunct}
{\mcitedefaultendpunct}{\mcitedefaultseppunct}\relax
\EndOfBibitem
\bibitem[Song and Chandler(1998)]{Song:1998}
X.~Song and D.~Chandler, \emph{The Journal of Chemical Physics}, 1998,
  \textbf{108}, 2594--2600\relax
\mciteBstWouldAddEndPuncttrue
\mciteSetBstMidEndSepPunct{\mcitedefaultmidpunct}
{\mcitedefaultendpunct}{\mcitedefaultseppunct}\relax
\EndOfBibitem
\end{mcitethebibliography}

\end{document}